\documentclass[prodmode,acmtist]{acmsmall} 

\usepackage[ruled]{algorithm2e}
\usepackage{tipa}
\usepackage{amsmath}
\usepackage{comment}

\SetAlFnt{\small}
\SetAlCapFnt{\small}
\SetAlCapNameFnt{\small}
\SetAlCapHSkip{0pt}
\IncMargin{-\parindent}

\newtheorem{mydef}{Definition}
\newtheorem{myproblem}{Problem}
\newtheorem{mysubproblem}{Subproblem}

\begin{document}

\markboth{}{}

\title{Unveiling Correlations
via Mining Human-Thing Interactions \\
in the Web of Things}
\author{Lina Yao
\affil{UNSW Australia}
Quan Z. Sheng
\affil{Macquarie University}
Anne H.H. Ngu
\affil{The Texas State University}
Xue Li
\affil{The University of Queensland}
Boualem Benattalah
\affil{UNSW Australia}
}

\begin{abstract}
With recent advances in radio-frequency identification (RFID), wireless sensor networks, and Web services, physical things are becoming an integral part of the emerging ubiquitous Web. Finding correlations among ubiquitous things is a crucial prerequisite for many important applications such as things search, discovery, classification, recommendation, and composition. This article 
presents {\bf DisCor-T}, a novel graph-based approach for discovering {\em underlying connections} of things via mining the rich content embodied in the human-thing interactions in terms of user, temporal and spatial information. We model these various information using two graphs, namely a spatio-temporal graph and a social graph. Then, random walk with restart (RWR) is applied to find proximities among things,
and a relational graph of things (RGT) indicating implicit correlations of things is learned. The correlation analysis lays a solid foundation contributing to improved effectiveness in things management and analytics. To demonstrate the utility of the proposed approach, we develop a flexible feature-based classification framework on top of RGT and perform a systematic case study. Our evaluation exhibits the strength and feasibility of 
the proposed approach.  

\end{abstract}




\acmformat{}

\begin{bottomstuff}
Lina Yao's research has been supported
by ARC Discovery Early Career Researcher Award DE160100509. Quan Z. Sheng's research has been partially supported by Australian Research Council (ARC) Future Fellowship
FT140101247 and Discovery Project Grant DP140100104. 
Authors’ addresses: L. Yao and B. Benatallah, School of Computer Science and Engineering, the University of New South Wales,
NSW 2052, Australia; email: \{lina.yao, boualem\}@cs.unsw.edu.au; 
Q. Z. Sheng, Department of Computing, Macquarie University, NSW 2109, Australia; email: michael.sheng@mq.edu.au; 
A. H. H. Ngu, Department
of Computer Science, Texas State University, TX 78666-4616, USA; email: angu@txstate.edu; 
X. Li, School
of ITEE, the University of Queensland, Queensland, 4072, Australia; email: xueli@itee.uq.edu.au.
\end{bottomstuff}

\maketitle

\section{Introduction}

Since its birth in early 1990s, the World Wide Web has been the heart of the research, development, and innovation in the world.
Indeed, it has changed our world and society so quickly and profoundly over the last two decades by sharing
knowledge and connecting people. Very recently, the World Wide Web is beginning to connect ordinary
things in the physical world, also 
called ``Web of Things'' (WoT)~\cite{christophe2011searching,guinard2011internet,Mathew2013,WoT-sheth2013,yao-IC2015}. 
As indicated by the inventor of the World Wide Web, Tim Berners-Lee, ``{\em it isn’t the documents which are actually
interesting; it is the things they are about!}''\footnote{\url{http://ercim-news.ercim.eu/en72/keynote/the-web-of-things}}.
WoT aims to connect everyday objects, such as coats, shoes, watches,
ovens, washing machines, bikes, cars, even humans, plants, animals,
and changing environments, to the Internet to enable communication/interactions between these objects. 
Being widely regarded as one of the most important technologies that is going to change our world in the coming decade, 
the ultimate goal of
WoT is to enable computers to see, hear and sense the real world. 


While such a ubiquitous WoT environment offers the capability of integrating information
from both the physical world and the virtual one leading to tremendous business and social opportunities (e.g., efficient supply chains, independent living of elderly persons, and improved environmental monitoring), it also presents significant challenges~\cite{Ferscha-2012,BaresiMD15,Vitali-2014,Sheng-Computer08}. With many things connected and interacted over the 
Web, there is an urgent need to efficiently index, organize, and manage these things for object search, recommendation, and mash-up, and effectively reveal interesting patterns from things. 

%

Before 
effectively and efficiently classifying, managing and 
recommending ubiquitous things, 
a fundamental 
task is to 
discover relations among things. 
Indeed, finding implicit correlations among things is a much more challenging task than finding relations for documents, web pages, and people, due to the following unique characteristics of things on the Web. 

\begin{itemize}

\item{{\bf Lack of uniform features}.} Things are diverse and heterogeneous in terms of functionality, access methods or descriptions. Some things have meaningful descriptions while many others do not~\cite{Search-WoT2011,linaicdm2013}. As a result, it is quite challenging to discover the implicit correlations among heterogeneous things. Things cannot be easily represented in a meaningful feature space. They usually only have very short textual descriptions and lack a uniform way of describing the properties and the services they offer~\cite{kindberg2002people}. 

\vspace{2mm}
\item{{\bf Lack of structural interconnections}.} Correlations among things are not obvious and are difficult to discover. Unlike social networks of people, where users have observable links and connections, things often exist in isolated settings and the explicit interconnections between them are typically limited. Such high level structural interconnection information
(e.g., a water tap and a cutting board are likely to be used together when cooking) are implicit in general \cite{linaicdm2013}. 

\vspace{2mm}
\item{{\bf Contextual uncertainty}.} 
Things are tightly bound to contextual information (e.g., location, time, status), as well as 
user behaviors (e.g., activities involving things), as things usually provide functionality-oriented services (e.g., washing vegetables for a water tap). Unfortunately, 
contextual information associated with things is highly dynamic (e.g., the location of a moving person changes all the time) and  
 has no obvious, easily indexable properties, which is unlike those 
static, human-readable texts in the case of documents \cite{guinard2011internet,yao2016things}. Capturing discriminative contextual information carried by 
things therefore is of paramount importance in effective things management. 
%
\end{itemize}

Some research efforts have proposed to explore things similarity and relations from semantic Web perspective \cite{mietz2013semantic,christophe2011searching}. In such cases, explicit relations of things can be characterized by using keyword-based, textual-level calculations. However, physical things also hold {\em implicit} relations due to their more distinctive structures and connections in terms of 
functionalities (i.e., usefulness), as well as non-functional attributes (i.e., availability). Different things provide different functionalities (e.g., microwave and printer), and might be of interest to different groups of people. With recent development of 
technologies such as radio frequency identification (RFID), wireless sensors, and Web services, 
human-thing interactions can be easily recorded and obtained (e.g., RFID readings). These interactions are not completely random. They carry rich information that can be harnessed and utilized to uncover the implicit relations. Although correlations between things are implicit, we argue that they can be captured by exploring regularities of user interactions with similar things.



This work targets mining 
useful information for unveiling implicit 
correlations of things from contextual information of human-thing interactions. Our proposed method, {\em DisCor-T} ([`disk\textschwa uti], {\bf dis}covering {\bf cor}relations of {\bf t}hings),
should be {\em effective} in capturing and reflecting the hidden structure of things from things usage events in the modeling stage, and {\em efficient} in inferring the related things in the inferring stage. 
Specifically, we present a novel approach that converts the things usage events into a {\em relational graph of things} (RGT)
by extracting three dimensional contextual information contained in the events history. The RGT graph underpins many important applications. 
We particularly present an application scenario to show its benefits in serving things clustering and annotation. To the best of our knowledge, no previous work has systematically studied mining the relationships of ubiquitous objects in WoT. The main contributions of our work can be summarized as follows:
\begin{itemize}
\item We study the problem of managing ubiquitous things 
 in the emerging Web of Things environment, 
which have unique characteristics (e.g., short descriptions, diverse, dynamic and noisy). 
We propose to investigate human-thing interactions from three contextual aspects: 
user, temporal, and spatial. Accordingly, we develop two graph presentations that approximate corresponding relationships from user-thing interactions. These graphs lay the foundation for uncovering latent correlations among things. 

\item We develop an algorithm for discovering latent correlations among things by applying Random Walk with Restart over the two contextual graphs. The learned correlations are used to construct the relational graph of things (RGT), which can help in a number of important applications on things management. In particular, we focus on a systematic case study on {\em things annotation}
to 
showcase the effectiveness of our approach.

\item We establish a testbed environment where things are tagged by RFID and sensors, and things usage events are collected in real-time. Using this real-world data with $\sim$20,000 records collected from the testing environment over a period of four months, we conduct extensive experimental studies to demonstrate the feasibility of our proposed approach.  
\end{itemize}

The remainder of the paper is organized as follows. In Section~\ref{sec:background}, we present some background information related to our work including motivating applications and 
formal definitions of the research problems. We then introduce the details of our proposed methodology {\em DisCor-T} in Section~\ref{sec:model}. 
We further demonstrate the benefits of our approach by designing a feature-based things annotation method in Section~\ref{sec:applications}. We report the implementation and experimental studies in Section~\ref{sec:evaluation}. Finally, we review the related work in Section~\ref{sec:relatedwork} and give some concluding remarks in Section~\ref{sec:conc}.

\section{Background}
\label{sec:background}
In this section, we first describe several application scenarios underpinned by the techniques discussed in this paper. We then formally formulate the research problems target by our work.

\subsection{Motivating Applications} 
Discovering underlying similarities except keyword-based similarity can allow for more meaningful and accurate things recommendation, classification and even contribute to context-aware activity recognition. We briefly discuss some of areas where things contextual similarity can be applicable. 

\begin{itemize}
\item {\bf Recommendation.} Things recommendation is a crucial step for promoting and taking full advantage of the Web of Things (WoT), where it benefits the individuals, businesses and society on a daily basis in terms of two main aspects. On the one hand, it can deliver relevant things to users based on users’ preferences and interests. On the other hand, it can also serve to optimize the time and cost of using WoT in a particular situation. 

The underlying correlations of things can enhance the performance of generalized recommendation systems in the Web of Things in terms of two main points. Firstly,
due to the sparsity of thing-user interactions, widely used collaborative filtering recommendation systems fail to find similar users or things, since these methods 
assume that two users have invoked at least some things in common. Moreover, users who have never used any things can not be fed good results in the first place (i.e., the cold start problem). 
Secondly, 
physical things have more distinctive structures and connections in terms of functionalities in real life (i.e., usefulness), as well as non-functionalities (i.e., availability), which are saliently highlighted in contextual information of human-thing interactions. 

\vspace{2mm}
\item{\bf Searching.} Developing efficient searching approaches is a crucial challenge 
with rapid increase of vast amount of things 
connected to the Web. Our approach adds one additional dimension to assist and reinforce 
current search techniques.
%
For instance, existing semantic-based solutions 
do not make full use of the rich information contained in users' historical interactions with things 
(e.g., 
implicit relations of different things). Our approach can effectively capture such information, which can be integrated 
into existing search solutions for better performance.
In particular, the latent connections between things/objects can be leveraged to predict which things 
might possibly co-occur for the search. 

\vspace{2mm}
\item{\bf Context-aware Activity Recognition.} Recognizing human activities from sensor readings has recently attracted much research interest in pervasive computing
due to its potential in many applications, such as assisted living of older people and healthcare. This task is particularly challenging because human
activities are often performed in a not only simple (i.e., sequential), but also complex (i.e., interleaved or concurrent) manner in real life.

Our proposed approach provides a new useful 
means to infer human activities by taking advantages of reasoning relationships of globally unique object instances. For example, dense sensing-based activity monitoring learns human activities by detecting and analyzing human-object interactions. By discovering correlations of objects, we can cluster and organize things into different structured groups based on their underlying relationships. In many cases, an activity could involve multiple relevant things including not only the things with similar functionalities but also things with complementary functionalities, which can be effectively uncovered by our proposed approach. 
Pairwise things with strong correlations indicate either they have similar functionalities (i.e., microwave and roaster) or they have 
a higher likelihood to be used together. For instance, a water tap and a chopping board are both in use when we prepare meals, since most of the time we need to wash cooking ingredients (e.g., vegetables) before cutting them.

\end{itemize}

\subsection{Problem Statement and Definitions}
\label{sec:problem}

The only data source used in our work is human-thing interactions, namely {\em things usage events}. Each event happens when a person interacts with a particular thing, which carries three kinds of information: location, timestamp, and user. Each usage event record can be defined as a quadruplet {\em ThingID, UserID, Timestamp, Location} described as follows. 

\begin{mydef}[Things Use Log]
\label{def:use}
Each thing use log happens when a person interacts with a particular thing. Let $\mathbf{O} = \{o_1,...,o_n\}$, $\mathbf{U} = \{u_1,...,u_m\}$, $\mathbf{Ts} = \{ts_1,...,ts_p\}$ and $\mathbf{Loc} = \{loc_1,...,loc_q\}$ represent the set of things, users, timestamps and locations, respectively. A usage event of a thing $o_i$, denoted by $h \in \mathbf{H} = \{h_1,...,h_i\} = \{<o,u,ts,loc> | o \in \mathbf{O} \land u \in \mathbf{U} \land ts \in \mathbf{Ts} \land loc \in \mathbf{Loc}\}$, indicates that user $u$ has used a particular thing $o$ located in a specific location $loc$. 
\end{mydef}

\begin{figure}[!tb]
\begin{center}
\begin{minipage}{6.5cm}
\includegraphics[width=6.5cm]{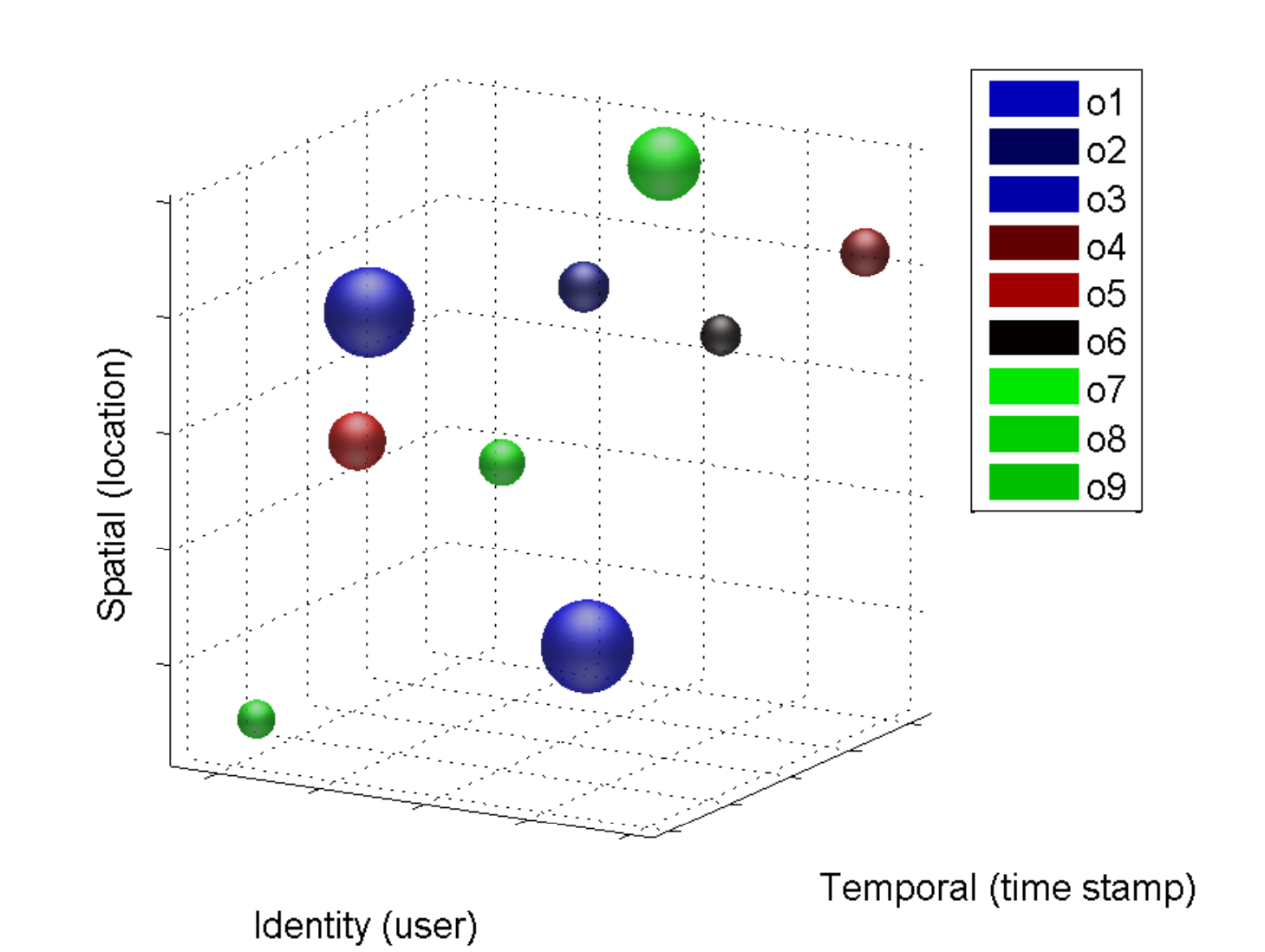}
\centering{(a)}
\end{minipage}
\begin{minipage}{6.5cm}
\includegraphics[width=6.5cm]{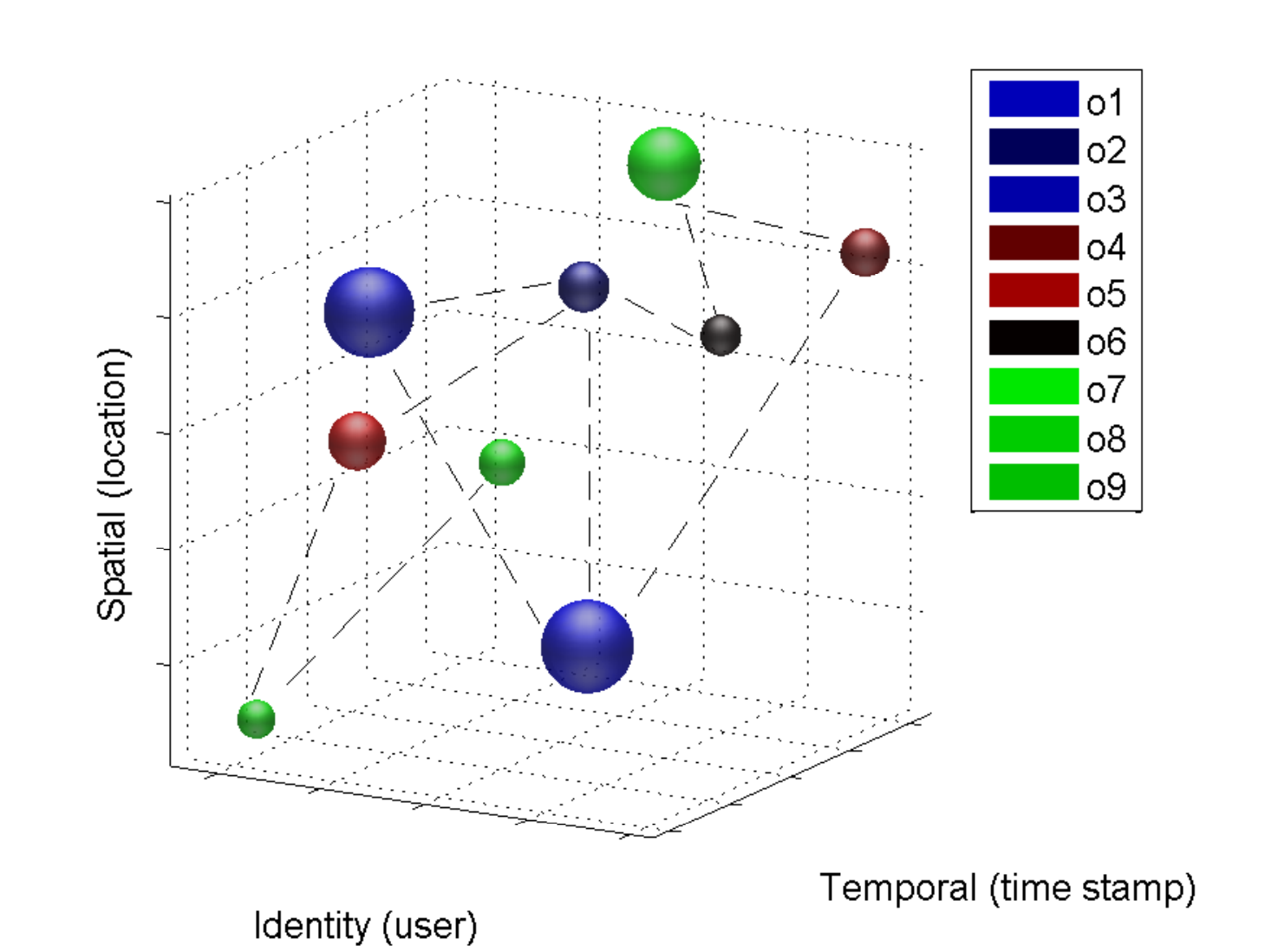}
\centering{(b)}
\end{minipage}
\caption{(a) Things usage and contextual attributes (b) Correlation discovery. Different colors of balls denote different things/objects, and the size represents the usage frequency.}
\label{fig:illustration}
\end{center}
\end{figure}

%
The problem targeted in this article can be therefore formulated as discovering the {\em latent correlations} among  things by exploiting observable {\em human-thing interactions} with the goal of automatically distinguishing strong correlations of things from the weak ones. As illustrated in Figure~\ref{fig:illustration}, each 
node denotes a thing (represented as a ball) in 
a three-dimensional space of identity, spatiality, and temporality. Things are discrete without distinctive and  explicit correlations (Figure~\ref{fig:illustration} (a)). However, our proposed approach can derive latent connections among these things and form a relational graph of things, where their implicit relatedness can be revealed (Figure~\ref{fig:illustration} (b)). Therefore, our goal can be formulated as follows in Problem~\ref{problem:overall}. 

\begin{myproblem}[Things Implicit Correlation Discovery]
\label{problem:overall}
\textit{
Given a 
set of human-thing interactions of quadruplets (thing, user, timestamp and location), discovering the latent 
correlations between things.}
\end{myproblem}

To complete this goal, 
there are two sequential subproblems we need to solve, which are defined as subproblem~\ref{subproblem:model} and subproblem~\ref{subproblem:infer} respectively. 

\begin{mysubproblem}[Modeling]
Given a collection of things usage events $\mathcal{H}$, construct two graphical models $\mathbf{G}_m$ capturing relations between things and their spatial-temporal information, and $\mathbf{G}_u$ capturing relations between things and users. 
\label{subproblem:model}
\end{mysubproblem}

\begin{mysubproblem}[Inferring]
Given the constructed graphs induced from things usage events collection $\mathbf{H}$, infer the similarities 
among things. 
\label{subproblem:infer}
\end{mysubproblem}


\section{Proposed Methodology}
\label{sec:model}

\begin{figure}[!tb]
\includegraphics[width=\linewidth]{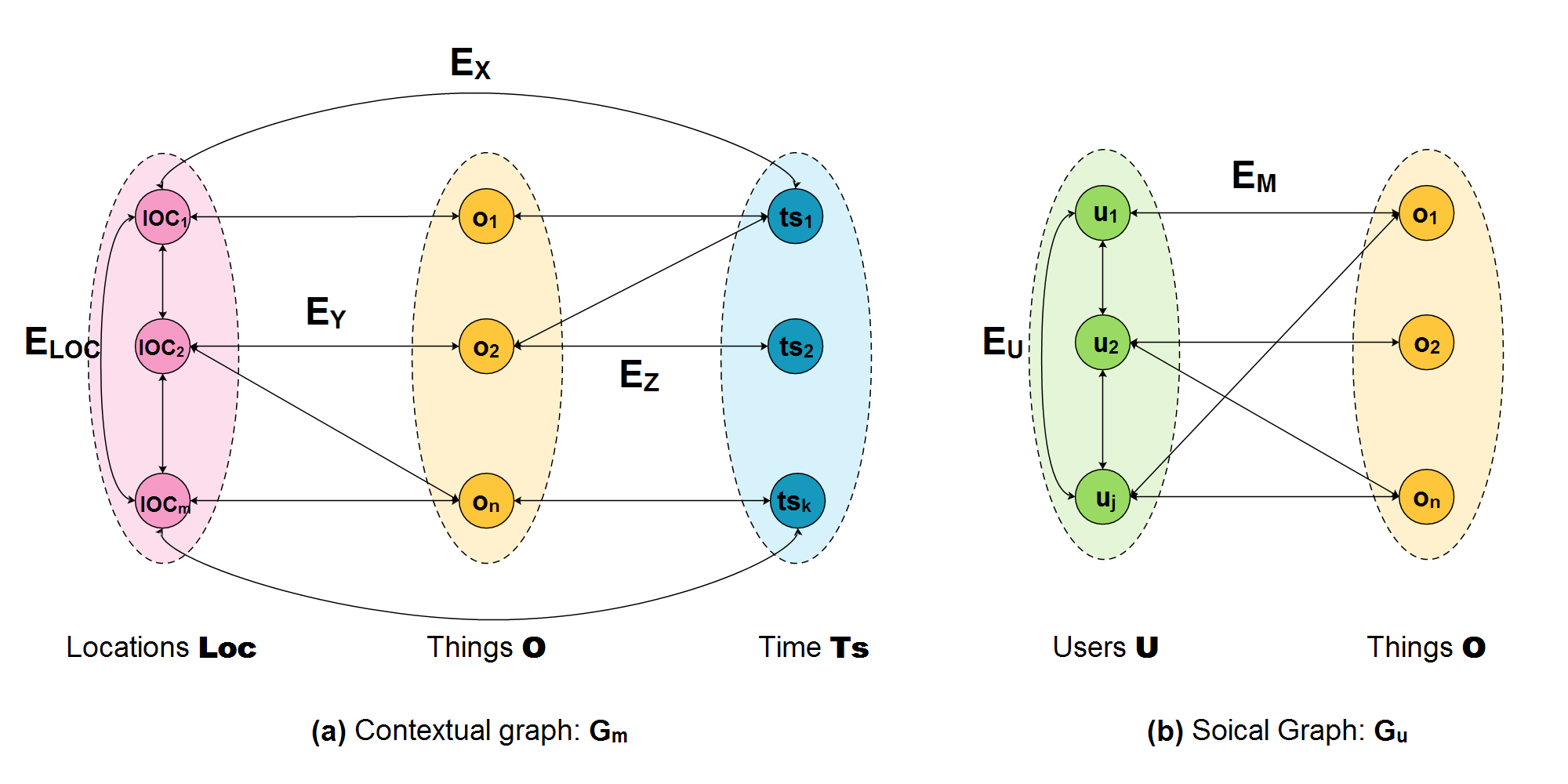}
\caption{Two graphs induced from things usage events: (a) spatio-temporal graph $\mathbf{G}_m$ (b) social graph $\mathbf{G}_u$.}
\label{fig:2graph}
\end{figure}

Our approach for correlation discovery of things involves two main stages corresponding to two subproblems defined in 
Section~\ref{sec:problem}.
The overall algorithm is shown in Algorithm~\ref{algo:correlation}. We firstly extract two types of graphs, namely 
the {\em location-time-thing} graph (Figure~\ref{fig:2graph}(a)) and the {\em user-thing} graph (Figure~\ref{fig:2graph}(b)). The graphs are deduced from thing usage events, which reflect object and its three related information in terms of 
spatio-temporal and social aspects.
Then we perform random walk on these two graphs respectively to inference relationships of pairwise things, and sum them up as the overall pairwise correlations of things. 

The first stage centers around building two graphs from things usage events. 
As illustrated in Figure~\ref{fig:2graph},  the {\em spatio-temporal} graph in Figure~\ref{fig:2graph} (a) 
captures the relations between things and their temporal and geographical influence, while the {\em social} graph in Figure~\ref{fig:2graph} (b) 
captures the social influence among users on interacting things. The technical details on how to construct these graphs will be described in Section~\ref{sec:context} and Section~\ref{sec:social} respectively.
In the second stage, our goal is to derive the pairwise relevance scores for things. To achieve this, a random walk with restart (RWR)~\cite{xia2009bi} is performed on the two 
constructed graphs. 
A relevance score is produced for any given node to any other node in the graph, 
presented as a converged probability. The value of the relevance score reflects the correlation strength between a pair of things. Based on the relevance scores, a top-$k$ {\em correlation graph of things} can be constructed, upon which many advanced things management problems such as annotation and clustering can be solved by tapping the wealth of literature in graph algorithms.  
The technical details on this part can be found in Section~\ref{sec:rwr}.

\begin{algorithm}[!tb]
\label{algo:correlation}
\caption{{\em DisCor-T}}
\SetAlgoLined
\KwIn{Sequences of things usage events $\mathbf{H}$ (Definition~\ref{def:use}), User friendship matrix $\mathbf{F}_{u}$}
\KwOut{Correlation matrix of things $\mathbf{R}$}
\vspace{0.1cm}
\%\%\% Stage 1: Graphs Construction (Section~\ref{sec:context} and Section~\ref{sec:social}) \%\%\%  \\
/*Detecting periodical connection between time and location*/ \\
\For{each location $loc_i \in \mathbf{Loc}$}{
Finding time periods for $l_i$ and store as $p_i$\;
Constructing edges between $l_i$ and $p_i$\;
}
Constructing 
spatio-temporal graph $\mathbf{G}_{m}$; \\
Constructing social graph $\mathbf{G}_{u}$; \\
\vspace{0.1cm}
\%\%\% Stage 2: Inferencing correlations via traversing graphs $\mathbf{G}_{m}$ and $\mathbf{G}_{u}$ (Section~\ref{sec:rwr}) \%\%\% \\
Obtaining transition probability matrix $\mathbf{P}_{m}$ and $\mathbf{P}_{u}$ deduced from corresponding weight matrix $\mathbf{W}_{m}$ and $\mathbf{W}_{u}$ respectively; \\
Implementing Random Walk with Restart (RWR) over $\mathbf{G}_{m}$ and $\mathbf{G}_{u}$ to derive things correlation matrix $\mathbf{R}_{m}$ and $\mathbf{R}_{u}$ respectively\;
Calculating weighted linear combination correlation matrix of things $\mathbf{R} = \alpha \mathbf{R}_{m}+ \beta \mathbf{R}_{u}$.

\end{algorithm}

%

\subsection{Spatio-Temporal Graph Construction}
\label{sec:context}

A spatio-temporal graph such as the one shown in Figure~\ref{fig:2graph}(a) 
reflects the temporal pattern and spatial information hidden in the things usage events. 
In our approach, the spatial and temporal information of things usage events is treated as 
inseparable
since they are mutually influential on detecting the correlations among things. 
Unlike virtual resources such as web pages, music or images, physical 
things such as restaurants and cookware usually provide 
more distinguished functionalists, and 
are more connected with people's daily lives. 
Some such distinctive features of physical things 
are their physical locations and 
functioning times. 
For example, kitchenware are more frequently used during dining times and they have 
higher likelihood to 
stay in a kitchen or similar locations (e.g., a dinning room). We specifically explore the integrity between spatial and temporal information in the ubiquitous things environment via finding the periodical pattern between time and locations. 

Generally, the timing of access of similar things may be similar. For example, restaurants are likely to be visited by people during lunch or dinner times. For the spatial information, we also argue in this paper that {\em geographical influence} to user activities cannot be ignored, i.e., a user tends to 
interact with the nearby things rather than 
the distant ones~\cite{ye2011semantic}.
For example, if a user is at her office, she has a higher probability of using office facilities such as telephone, desktop computer, printer, and seminar rooms. 


A spatio-temporal graph has three
sets of 
nodes,
namely locations, things, and timestamps.
It contains
one type of intra-relation (i.e., representing similarities between locations)
and three types of inter-relations between locations, things, and timestamps.
Edges between times and things can be obtained from usage events, say, the weight of edge $<loc,o> \in \mathbf{E}_{\mathbf{Y}}$ and $<ts,o> \in \mathbf{E}_{\mathbf{Z}} $ is proportional to the number of times objects $o$ is used in a location $loc$ and at timestamp $ts$. 
The inter-relation between location and time $<loc,ts> \in \mathbf{E}_{\mathbf{X}}$, indicates the periodical patterns. 
%
Formally, we define the spatio-temporal graph $\mathbf{G}_{m}$ as the following:

\begin{mydef}[Spatio-Temporal Graph]
A spatio-temporal graph is denoted by $\mathbf{G}_{m} =
<\mathbf{V}_m, \mathbf{E}_m>$. Here $\mathbf{V}_m$ = $\mathbf{Loc}\cup\mathbf{Ts}\cup \mathbf{O}$ where $\mathbf{Loc}$, $\mathbf{Ts}$ and $\mathbf{O}$ are the sets of locations, timestamps and things respectively. Edges $\mathbf{E}_{m} = \mathbf{E}_{\mathbf{Loc}}\cup \mathbf{E}_{\mathbf{X}}\cup \mathbf{E}_{\mathbf{Y}} \cup \mathbf{E}_{\mathbf{Z}}$, where $\mathbf{E}_{\mathbf{Loc}} = \{(loc,loc'):(loc, loc') \in \mathbf{Loc}\times \mathbf{Loc}\}$ and the weight of each edge $E_{loc}(i,i') \in \mathbf{E}_{\mathbf{Loc}}$ is associated with the similarity between location $i$ and $i'$. $\mathbf{E}_{\mathbf{X}} = \{(loc, ts):(loc, ts) \in \mathbf{Loc}\times \mathbf{Ts}\}$ and the weight of each edge $E_{X}(i,j) \in \mathbf{E}_{\mathbf{X}}$ is associated with a binary value, referring to whether location $loc_i$ has periodic relationship with time interval $ts_j$. $\mathbf{E}_{\mathbf{Y}} = \{(loc, o):(loc, o) \in \mathbf{Loc}\times \mathbf{O} \}$ and the weight of each edge $E_{Y}(i,j) \in \mathbf{E}_{\mathbf{Y}}$ is associated with the frequency that thing $o_j$ in location $loc_i$ is accessed. $\mathbf{E}_{\mathbf{Z}} = \{(ts, o):(ts, o) \in \mathbf{Ts}\times \mathbf{O} \}$ and the weight of each edge $E_{Z}(i,j) \in \mathbf{E}_{\mathbf{Z}}$ is associated with the frequency that thing $o_j$ is accessed in time interval $ts_i$.
\end{mydef}

The corresponding weight matrix $\mathbf{W}_m$ of graph $\mathbf{G}_{m}$ can be formulated as:
\begin{equation}
\mathbf{W}_m = 
\begin{bmatrix}
\mathbf{W}_{\mathbf{Loc}} & \mathbf{{W}_{X}} & \mathbf{{W}_{Y}}\\
\mathbf{{W}_{X}^{T}} & \mathbf{{W}_{Ts}} & \mathbf{{W}_{Z}}\\
\mathbf{{W}_{Y}^{T}} & \mathbf{{W}_{Z}^{T}} & \mathbf{{W}_{O}} \\
\end{bmatrix}
\label{mat:weight}
\end{equation}
where each of the entries in Equation~\ref{mat:weight} can be obtained as the following. $\mathbf{{W}_{Loc}}$ indicates the similarity of each pair of locations. Given two locations, we measure their similarity using the Jaccard coefficient between the sets of things used at each location: 
\begin{equation}
\mathbf{{W}_{Loc}}(i,j) = \dfrac{|\Gamma^{o}_{i}\cap\Gamma^{o}_{j}|}{|\Gamma^{o}_{i}\cup\Gamma^{o}_{j}|} 
\end{equation}
where $\Gamma^{o}_{i}$ and $\Gamma^{o}_{j}$ denote the set of used things at location $i$ and location $j$ respectively. 
$\mathbf{{W}_{Ts}}$ 
and $\mathbf{{W}_{O}}$
should be 0 since we do not consider the relationships between timestamps and the ones between things. 
$\mathbf{{W}_{Y}}$ and its transpose $\mathbf{{W}_{Y}^{T}}$ are integers, 
indicating how often a thing is 
accessed at a location.  Similarly, ${\mathbf{W_{Z}}}$ and its transpose $\mathbf{{W}_{Z}^{T}}$ are integers, which indicate how often a thing is accessed at a 
particular time.

For defining relationship between time stamps and locations and their corresponding weight $\mathbf{{W}_{X}}$ of graph $\mathbf{G}_{m}$, we propose 
{\em periodic patterns} between locations and timestamps. 
Given a sequence of locations $Loc = \{loc_1,...,loc_n\}$, our aim is to find their corresponding time period. To obtain relationship between time and location, 
we 
analyze the potential periods for each location and 
find the periodical pattern between locations and timestamps. 
%
A periodic pattern represents 
the repeat of certain usage event at a specific location with 
certain time interval(s).

\begin{figure}[!tb]
\begin{center}
\begin{minipage}{6.5cm}
\includegraphics[width=6.5cm]{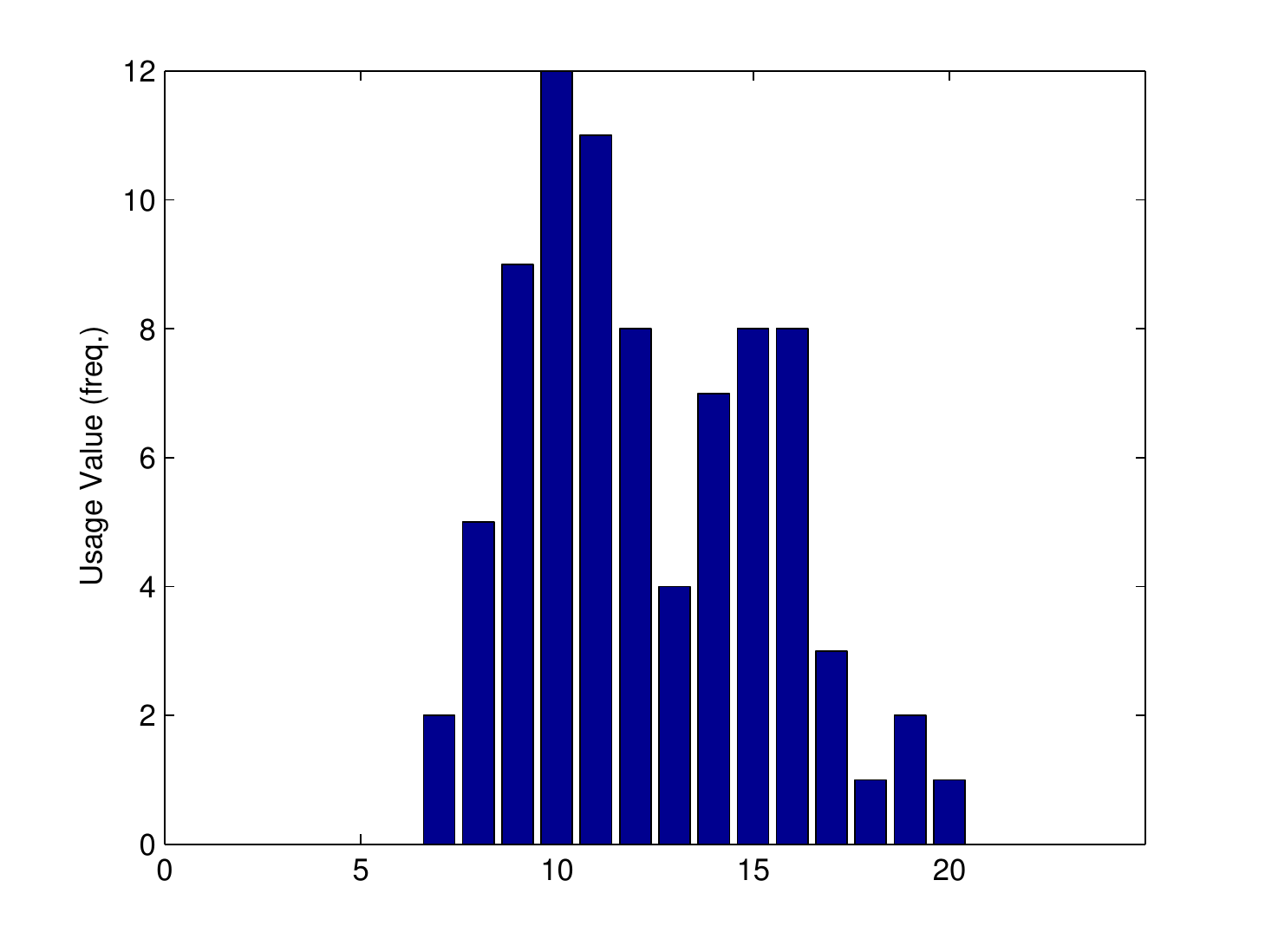}
\centering{(a)}
\end{minipage}
\hspace{5mm}
\begin{minipage}{6cm}
\includegraphics[width=6cm]{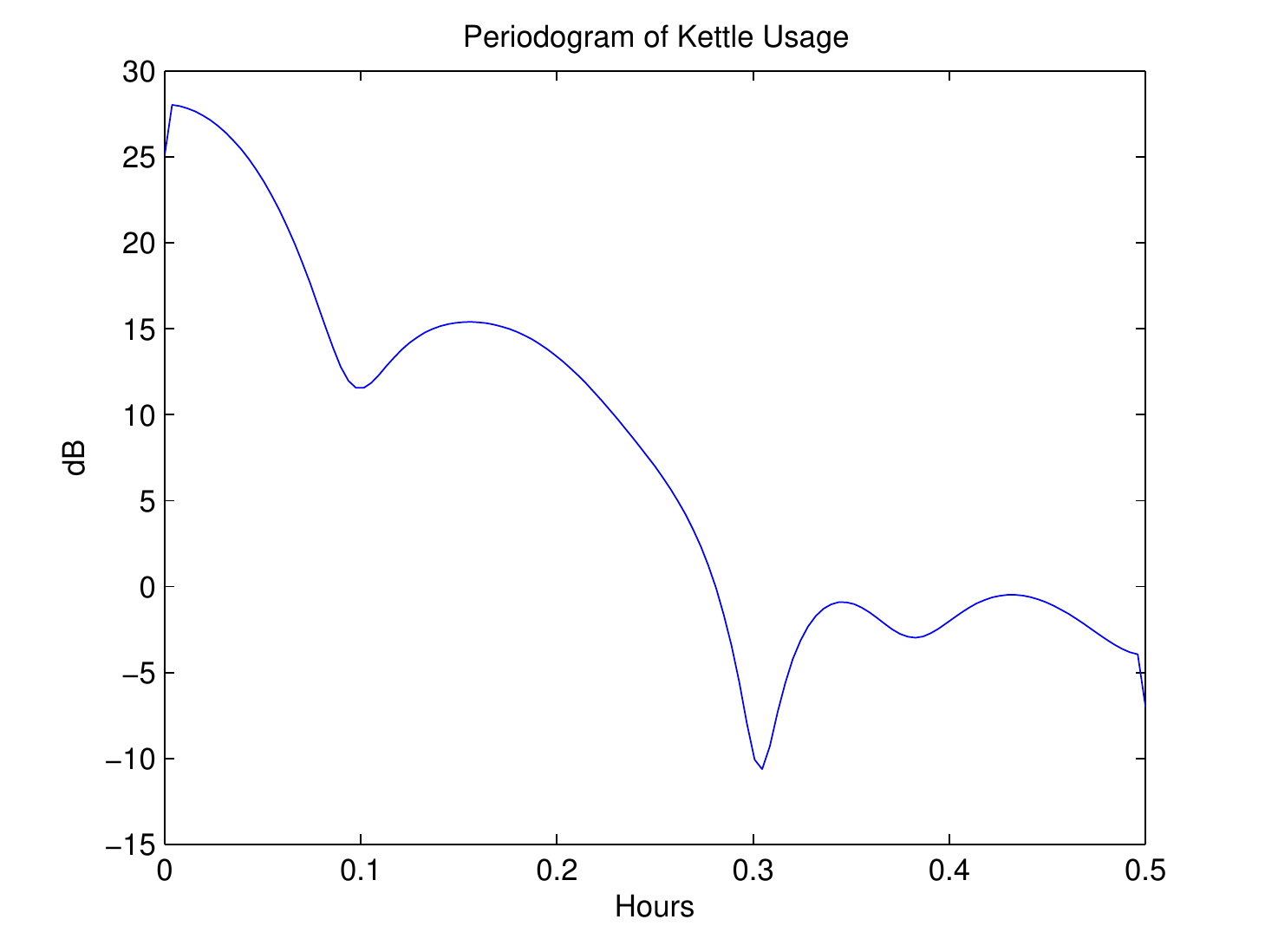}
\centering{(b)}
\end{minipage}
\caption{(a) Usage history of a kettle in the kitchen area over 24 hours within one week (e.g., the kettle was used 12 times at 10am); (b) Periodogram of the kettle usage}
\label{fig:usageexample}
\vspace{-4mm}
\end{center}
\end{figure}

Periodic patterns can be extracted by analyzing things usage events. In particular, we build a time series dataset for each location where the elements of the time series data are the number of time slots (e.g., 0 for the period of 0:00-1:00; 1 for 1:00-2:00 and so on) that a thing at a location is invoked. We can clearly observe such periodic pattern from the example relating to a kettle in the kitchen from Figure \ref{fig:usageexample} (a) and its periodogram in Figure \ref{fig:usageexample} (b).

Given a sequence of locations, 
we adopt the Discrete Fourier Transform (DFT) method to detect the time periods in this discrete time-series sequence \cite{vlachos2004identifying}. 
For each location, we define an integer sequence $A = \{a_{1}a_{2}...a_{n}\}$, where $a_i$=1 if the thing is used at this location at time $i$, and $0$ otherwise. Essentially, this sequence can be transformed into a sequence of $n$ complex numbers $X(f)$ from the time domain to the frequency domain:
\begin{equation}
X(f_{k/N}) = \dfrac{1}{\sqrt{N}}\sum_{n=0}^{N-1}a_{n}e^{-\dfrac{j2\pi kn}{N}}, k=0,...,N-1
\end{equation}
where $k/N$ denotes the frequency that each coefficient captures. As a result, DFT transforms the original sequences as a linear combination of the complex sinusoids $s_{f}(n) = \dfrac{e^{j2\pi fn/N}}{\sqrt{N}}$. 
The Fourier coefficients represent the amplitude of each of these sinusoids after sequences $S$ is projected on them. 

We aim at capturing the general shape of time-series data (e.g., thing usage over time) as ``economically'' as possible. 
To do so, we propose to use a 
spartan representation\footnote{Inspired by the frugal lifestyle of the ancient Spartans, a spartan representation means an economic way of representing a dataset in a smaller size~\cite{hristopulos2003spartan}.} from which one could reconstruct the signal using just its dominant frequencies (i.e., the ones that carry most of the signal energy). 
A popular way to identify the power content of each frequency is by calculating the {\em power spectral density} (PSD) 
of the sequence which indicates the signal power at each frequency in the spectrum. A well known estimator of the PSD is the {\em periodogram}. The periodogram $P$ is a vector comprised of the squared magnitude of the Fourier coefficients $X(f_{k/N})$:
\begin{equation}
P(f_{k/N}) = ||X(f_{k/N})||^2, k=0,1,...,\left \lceil \dfrac{N-1}{2} \right \rceil
\end{equation}
The $k$ dominant frequencies appear as peaks in the periodogram (and correspond to the coefficients with the highest magnitude). In order to specify which frequencies are important, we need to set a threshold and identify those frequencies higher than this threshold. Each  element  of the  periodogram provides  the power at frequency $k/N$ or, equivalently, at period $N/k$. 
That is, coefficient $X(f_{k/N})$ corresponds to periods $[\dfrac{N}{k},...,\dfrac{N}{k-1})$. Interested readers are referred to \cite{vlachos2004identifying}.  

When obtaining the periodgram of each location, we can decide their corresponding peak points based on 
preset threshold. From the periodgram, we can find the location and its corresponding time range. One benefit of using the periodogram is that we can visually identify the peaks as the $k$ most dominant periods (period =1/frequency). 
For automatically returning the important periods for a set of location sequences,
we can simply set a threshold in the power spectrum to distinguish the dominant periods. 
%
%
%
%
In Section~\ref{sec:data}, we describe how to extract location and time relationship from the usage events.  

\subsection{Social Graph Construction}
\label{sec:social}
Users' relations (e.g., friendships) can have significant impact on things usage patterns. Personal tastes are correlated. Research in \cite{kameda1997centrality} shows that friendships and relations between users play a substantial role in human decision making in social networks. For instance, people usually turn to a friend's advice about a commodity
(e.g., hair straighter) or a restaurant before they go for them. 
For exploring the impact social links between users on things' correlation discovery, we also construct a social graph, which is an augmented bipartite graph representing user interactions with things based on things usage events. 
As shown in Figure~\ref{fig:2graph} (b), such a graph 
contains two sets of entities, 
users $\mathbf{U}$ and things $\mathbf{O}$.
There is one type of intra-relation 
between users (also called social connections)
and one type of inter-relations: edges between users and things that can be obtained from usage events. Formally, the social graph is defined as the following:

\begin{mydef}[Social Graph]
A social graph, denoted by $\mathbf{G}_{u} = <\mathbf{V}_u, \mathbf{E}_u>$, is an augmented undirected bipartite graph. Here $\mathbf{V}_u$ = $\mathbf{U}\cup\mathbf{O}$ where $\mathbf{U}$, $\mathbf{O}$ are the sets of users and things respectively. Edges $\mathbf{E}_{u} = \mathbf{E}_{\mathbf{U}}\cup \mathbf{E}_{\mathbf{M}}$, where
$\mathbf{{E}_{U}}= \{(u,u'):(u,u') \in \mathbf{U}\times \mathbf{U}\}$ denotes the user social links (friendship) and each edge $E_{U}(i,i') \in \mathbf{E}_{\mathbf{M}}$ is associated with the similarity between user $u_i$ and user $u_i'$. $\mathbf{E}_{\mathbf{M}} = \{(u,o):(u,o) \in \mathbf{U} \times \mathbf{O}\}$. In this graph, each edge between users and things $E_{M}(i,j) \in \mathbf{E}_{\mathbf{M}}$ is associated with the frequency that thing $o_j$ is accessed by user $u_i$.
\end{mydef}

The corresponding weight matrix $\mathbf{W}_u$ of graph $\mathbf{G}_{u}$ can be formulated as:
\begin{equation}
\mathbf{W}_u = 
\begin{bmatrix}
\mathbf{W}_{U} & \mathbf{W}_{M} \\
\mathbf{W}_{M}^{T} & \mathbf{W}_{O}\\
\end{bmatrix}
\label{mat:weightuser}
\end{equation}

The entries in Equation~\ref{mat:weightuser} can be obtained as follows: $\mathbf{W}_{M}$ and its transpose $\mathbf{W}_{M}^{T}$ should be proportional to the number of times of a thing being used by the users. 
$\mathbf{W}_{O}$ should be zero since we do not consider relationships between things.
The weight $\mathbf{W}_{U}$ of edges $\mathbf{E}_{M}$ indicates the user similarity influenced by the social links between users, reflecting the homophily meaning that similar users may have similar interests. We use the cosine similarity to calculate $\mathbf{W}_{U}$ as follows:
\begin{equation}
\mathbf{W}_{\mathbf{U}}(i,j) = \dfrac{e^{\alpha cos(b(i),b(j))}}{\sum_{k\in \Omega(i)}e^{\alpha cos(b(i),(b(k))}}
\end{equation}
where $cos(b(i),b(j))= \dfrac{b(i)\cdot b(j)}{||b(i)||||b(j)||}$, $\Omega(i)$ is the set of the user $i$'s friends (i.e., $j \in \Omega(i)$), $b(i)$ is the binary vector of things used by user $i$, $||\cdot||$ is the L-2 norm of vector $b(\cdot)$, and $\alpha$ is a parameter that reflects the preference for transitioning to a user who interacted with the same things.

\subsection{Correlation Inference}
\label{sec:rwr}

After the two graphs $\mathbf{G}_{m}$ and $\mathbf{G}_{u}$ are constructed, 
we can perform the random walk with restart (RWR)~\cite{xia2009bi} to 
derive the correlation between each pair of things. 
RWR provides a good relevance score between two nodes in a graph, and has been successfully used in many applications such as automatic image captioning, recommendation systems, and link prediction. The goal of using RWR in our work is to find other things that have top-$k$ highest relevance scores for a given thing. The values of the relevance scores imply the strength of the correlations among things. In the following, we focus on using RWR on the spatio-temporal graph $\mathbf{G}_{m}$ for discovering correlations between things.

We assume the random walker starts from a thing node $o_i$ on $\mathbf{G}_m$. 
The random walker iteratively transits to other nodes which have edges with $o_i$, with the probability proportional to the edge weight between them. At each step, 
$o_i$ also has a restart probability $c$ to return to itself. We can obtain the steady-state probability of 
$o_i$ visiting other vertex $\pi_i$ when the RWR process is converged. The RWR process can be formulated as
\begin{equation}
\pi_i = (1-c)\mathbf{P}\pi_i + c\mathbf{e}_i
\label{equ:initialrwr}
\end{equation}
where $\pi_i \in \mathbb{R}^{N \times 1}$, and weight matrix from graph $\mathbf{G}_m$ is $\mathbf{W}_m \in \mathbb{R}^{N \times N}$(Section~\ref{sec:context}), $\mathbf{e}_i \in \mathbb{R}^{N \times 1}$ with $i$-th entry is 1, all other entries are 0. 
Equation~\ref{equ:initialrwr} can be further formulated as:
\begin{equation}
\pi_i = c(\mathbf{I}-(1-c)\mathbf{P}_m)^{-1}\mathbf{e}_i = \mathbf{Q}\mathbf{e}_i
\label{equ:q}
\end{equation}
where $\mathbf{I}$ is an identity matrix and $\mathbf{P}_m \in \mathbb{R}^{N \times N}$ is the transition matrix, which can be obtained based on weight matrix $\mathbf{W}_m$ of $\mathbf{G}_m$ by  row normalization:
\begin{equation}
\mathbf{P}_m = \mathbf{W}_m\mathbf{D}_{m}^{-1}
\label{equ:transition}
\end{equation}
where $\mathbf{D}_m$ is a diagonal matrix with $\mathbf{D}_{m}(i,i)=\sum_{j}\mathbf{W}_{m}(i,j)$. The random walker  on thing $o_i$ traverses randomly along its edges to the neighboring nodes based on the transition probability $\mathbf{P}_{m}(i,j), \forall j \in N(i)$, and the probability of taking a particular edge $<$$o_i$,$o_j$$>$ is proportional to the edge weight over all the outgoing edges from $o_i$ based on Equation~\ref{equ:transition}. 

In Equation~\ref{equ:q}, $\mathbf{Q} = c(\mathbf{I} - (1-c)\mathbf{P}_m)^{-1} = c\sum_{t=0}^{\infty}(1-c)^{t}\mathbf{P}^{t}$ defines all the steady-state probabilities of random walk with restart.
$\mathbf{P}^{t}$ is the $t$-th order transition matrix, whose elements $p_{ij}^{t}$ can be interpreted as the total probability for a random walker that begins at node $i$ and ends at node $j$ after $t$ iterations, considering all possible paths between $i$ and $j$. 
Since in our case we only consider relevance score between two things, 
if we vary the value of $t$, we can explicitly explore relationship between two things at different scales. 
The steady-state probabilities for each pair of nodes can be obtained by recursively processing Random Walk and Restart until convergence. The converged probabilities give us the long-term visiting rates from any given node to any other node. 
This way, we can obtain the relevance scores of all pairs of 
thing nodes, denoted by $R_m(o_i,o_j) \in \mathbf{R}_m, \forall o_i,o_j \in \mathbf{O}$. 
It should be noted that the results can be calculated more efficiently by using the Fast Random Walk with Restart implementation~\cite{tong2006fast} via low-rank approximation and graph partition.

Similarly, the transition probability matrix $\mathbf{P}_u$ for the social graph $\mathbf{G}_u$ can be obtained using: 
\begin{equation}
\mathbf{P}_u = \mathbf{W}_{u}\mathbf{D}_{u}^{-1}
\end{equation} 
where $\mathbf{D}_u$ is a diagonal matrix with $D_{u}(i,i)=\sum_{j}\mathbf{W}_{u}(i,j)$. Accordingly, we can obtain the relevance scores of things on this graph $R_u(o_i,o_j) \in \mathbf{R}_u, \forall o_i,o_j \in \mathbf{O}$. 

The overall relevance score (i.e., the correlation value) of any pair of things can be calculated using
\begin{equation}
R(o_i,o_j) = \alpha R_m(o_i,o_j) + \beta R_u(o_i,o_j)
\label{equ:overall}
\end{equation}
where $\alpha \in [0,1]$ and $\beta \in [0,1]$, which are regulatory factors affecting the weight on the social influence and the spatio-temporal influence.

With obtained correlation values, we could construct a top-$k$ {\em correlation graph of things}
by connecting each thing 
with the things that have top-$k$ overall correlation values $R(o_i,o_j)$. 
Formally, the graph is defined as the following:

\begin{mydef}[Relational Graph of Things (RGT)]
RGT is denoted by $\mathbf{G} = (\mathbf{O}, \mathbf{E})$. For each thing $o_i \in \mathbf O$, 
let $\mathbf{O}_i^{k}$ denote the top-$k$ set of correlative things to $o_i$. $\mathbf E = \{e(x, i)|\forall o_i \in \mathbf{T}, o_x \in \mathbf{O}_i^{k} \}$, where $e(x,i)$ is an edge from $o_x$ to $o_i$. Each edge is associated with a 
weight $w_{o_x, o_i}$
with the correlation value $R_{o_x,o_i}$.
\end{mydef}

\section{Applicability of DisCor-T: Things Classification} 
\label{sec:applications}

The top-$k$ correlation graph $\mathbf{G}$ is essentially a {\em graph} representing the relationships among things.
For instance, from our experiment, we found that the top four things most close to a three-seated sofa are modular sofa,
leather chair,
high chair,
and wooden chair.
Using the constructed $\mathbf{G}$, many problems centered around things management (e.g., things discovery, search and recommendation) can be solved and 
explored further by exploiting existing graph algorithms. 
In this section, we will showcase the feasibility and effectiveness of our proposed DisCor-T by detailing one important research problem, {\em automatic things annotation}, which will be used later to evaluate the performance of our proposed approach to correlation discovery.

Automatically predicting appropriate tags (i.e., category labels) for 
unlabeled things can save manual labeling workload, and has important research significance. 
Although some things have been labeled with useful tags (e.g., \texttt{cooking}, \texttt{office}),
which are crucial for assisting users in searching and exploring new things, as well as recommending them, some other things may not have any meaningful labels at all.
Furthermore, a thing might be associated with multiple categories.  
For instance, a microwave oven can be categorized in \texttt{Cooking} and also  \texttt{Home Appliance}.
%

The aim of things annotation is that when given a new thing, 
the classifier automatically decides whether this thing belongs to the category of the corresponding labels. The algorithm can be divided into two stages: 
i) extracting features from the
top-$k$ correlation graph $\mathbf{G}$ and things, and ii) 
performing multi-label classification of things. We extract three kinds of features $\mathbf{F_L}$, $\mathbf{F_S}$ and $\mathbf{F_C}$ from RGT in terms of label property, link structures and node attributes respectively. 

\paragraph{Extracting feature $\mathbf{F_L}$} This feature 
represents the label probabilities for 
unknown things, 
%
which can be 
computed using generative Bayesian rules from $\mathbf{G}$, where each 
unknown thing $o^*$ is to be assigned one or multiple labels $l_k \in \mathbf{L} = \{l_1,...,l_k\}$. 
We 
propose to formulate our solution as posterior probability $Pr(l_k|o^*)$. Once we know these probabilities, it is straightforward to assign $o_i$ the label having the top-$K$ largest probabilities, 

\begin{equation}
\begin{split}
Pr(l_k|o^{*}) &= \dfrac{Pr(o^{*}|l_k)Pr(l_k)}{\sum_{j=1}^{K}Pr(x|l_j)Pr(l_j)} \propto Pr(o^{*}|l_k)p(l_k)
\end{split}
\end{equation}
where the prior distribution probability $Pr(l_k)$ can be 
easily 
calculated from the training dataset. 
Let $\mathbf{o}^{l_{k}} = o_{1}^{l_k},...,o_{M_k}^{l_k}$ be the training dataset, having $M_k$ things with label $k$. 
Then $Pr(o^{*}|l_k)$ can be calculated using:
\begin{equation}
\begin{split}
Pr(o^{*}|l_k) &= \dfrac{1}{Z}\sum_{m=1}^{M_k}Pr(o^{*}|o_{m}^{l_k},l_k)Pr(o_m^{l_k}|l_k) \\
& = \dfrac{1}{Z\times M_k}\sum_{m=1}^{M_k}Pr(o^{*}|o_{m}^{l_k},l_k)
\end{split}
\label{equ:annotate}
\end{equation}
where $Z$ is a normalizing constant and the conditional probability $Pr(o^{*}|o_{m}^{l_k},l_k)$ indicates the relevance score between testing thing $o^{*}$ and things in the training dataset $o_{m}^{l_k}$. 
$Pr(o^{*}|o_{m}^{l_k},l_k) \approx \mathbf{\pi}_{o^*}$ denotes the steady state probability between $o^{*}$ and $\mathbf{o}^{l_{k}} = o_{1}^{l_k},...,o_{M_k}^{l_k}$, which can be obtained from Equation~\ref{equ:q} in our RWR process. 
The distribution $p(o_{m}^{l_k}|l_k)$ is set as a uniform distribution $1/M_k$. 
The probability $p(o^{*}|l_k)$ can be predicted in Equation~\ref{equ:annotate}, 
and the labels with different posterior probabilities can be assigned to the testing thing. 
As a result, we can get the label probabilities for each testing object. 
\paragraph{Extracting latent feature $\mathbf{F_S}$} 


With RGT, we can easily extract the features of things from RGT indicating the things relationship with different communities on $\mathbf{G}$. In reality, things usually hold multiple relations. For instance, a thing might be shared among its owner, owner's friends, co-workers, or family members. It might also be connected to other things based on functionality or non-functionality attributes. Detecting such relations from RGT, which can be used as a structural feature for things annotation, is naturally related to the task of {\em modularity-based} {\em community detection}~\cite{leicht2008community}. Modularity is to evaluate the goodness of a partition of undirected graphs. The reason that we choose this method is that modularity has been shown to be an effective quantity to measure community structure in many complex networks~\cite{tang2009relational}. 

Modularity $\mathcal{Q}$ is like a statistical test that the null model is a uniform random graph model, where
one vertex connects to others with uniform probability. It is a measure of how far the interaction deviates from a uniform random graph with the same degree distribution. Modularity is defined as:

\begin{equation}
\mathcal{Q} = \dfrac{1}{2m}\sum_{ij}\bigg[\mathcal A_{ij} - \dfrac{d_{i}d_{j}}{2m}\bigg] \delta(s_i,s_j)
\end{equation}

Where $\mathcal A_{ij}$ is the adjacent matrix on the graph RGT, $m$ is the number of edges of the matrix, $d_{i}$ and $d_{j}$ denote the in-degree of vertex $i$ and out-degree of vertex $j$, and $\delta(s_{t_i}, s_{t_j})$ are the Kronecker delta function that takes the value 1 if node $t_i$ and $t_j$ belong to the same community, 0 otherwise. A larger modularity $\mathcal{Q}$ indicates denser within-group interaction. So that, the modularity-based algorithm aims to find a community structure such that $\mathcal{Q}$ is maximized. In~\cite{newman2006finding}, Newman proposes an efficient solution by reformulating  $\mathcal{Q}$ as:

\begin{equation}
\mathcal{Q} = \dfrac{1}{2m}\mathcal S^{\mathcal T}\mathbf{B}\mathcal S
\end{equation}
where $\mathcal S$ is the binary matrix indicating which community each node belongs to. $\mathbf{B}$ is the modularity function, 
is defined as the following:
\begin{equation}
\mathcal B_{ij} = \mathcal A_{ij} - \dfrac{d_{i}d_{j}}{2m}
\end{equation}
Since our relational graph of things (RGT) is a weighted and directed graph, 
we need to make some modifications on $\mathcal{Q}$ to solve the equation. This involves two steps.

In the first step, we extend $\mathcal B$ to directed graphs.
Based on~\cite{leicht2008community}, we rewrite the modularity matrix $\mathcal {B}$ as the following:

\begin{equation}
\mathcal B'_{ij} = \mathcal A_{ij} - \dfrac{d_{i}^{in}d_{j}^{out}}{2m}
\label{eqn:bb}
\end{equation}
where ${d_{i}^{in}d_{j}^{out}}$ are the in-degrees and out-degrees of all the nodes on the RGT graph. In the second step, we extend $\mathcal B'$ to weighted graphs. 
To do so, we conduct further modification based on Equation~\ref{eqn:bb}. It can be rewritten further as below:

\begin{equation}
\mathcal B''_{ij} = \mathcal W_{ij} - \dfrac{w_{i}^{in}w_{j}^{out}}{2m}
\end{equation}
where $\mathcal W_{ij}$ is the sum of weights of all edges in the RGT graph replacing the adjacency matrix $\mathcal A$,     
 $w^{in}_{i}$ and $w^{out}_{j}$ are the sum of the weights of incoming edges adjacent to vertex $t_i$ and the outgoing edges adjacent to vertex $t_j$ on the RGT graph respectively. After these two steps, it should be noted that different from undirected situation, $\mathcal {B}''$ is not symmetric. To use the spectral optimization method proposed by Newman in~\cite{newman2006finding}, we restore symmetry by adding $\mathcal {B}"$ to its own transpose~\cite{leicht2008community}, thereby the new $\mathcal{Q}_{new}$ is:
\begin{equation}
\mathcal Q_{new} = \dfrac{1}{4m}\mathcal S^\mathcal T(\mathcal {B}''+\mathcal B''^{\mathcal T})\mathcal S
\end{equation}

We then is able to calculate all the eigenvectors corresponding to the top-$k$ positive eigenvalue of $\mathcal {B}''+\mathcal {B}''^{\mathcal T}$ and assign communities based on the elements of the eigenvector~\cite{newman2004finding}. We take the obtained modularity vectors as the latent features, which indicate things relationships to communities (i.e., a larger value means a closer relationship with a community).

\paragraph{Extracting feature $\mathbf{F_C}$} It is the set of content-based features extracted from thing descriptions. We convert the keywords vectors into tf-idf format, which assigns each term $x$ a weight in a thing's description $d$. $tf-idf(x,d) = tf(x,d)\times idf(x)$, where $tf(x,d)$, the number of times word $x$ occurs in the corresponding thing's description $d$, and $idf$ is the inverse text frequency which is defined as : $idf(x) = \log\dfrac{|N|}{df(x)}$, where $|N|$ is the number of texts in our dataset, and $df(x)$ is the number of texts where the word $x$ occurs at least once.

Based on our experience in ontology bootstrapping for Web services~\cite{Aviv-TSC2012}, we exploit Term Frequency/Inverse Document Frequency (TF/IDF)---a common method in IR for generating a robust set of representative keywords from a corpus of documents---to analyze things' descriptions. It should be noted that the common implementation of TF/IDF gives equal weights to the term frequency and inverse document frequency (i.e., $w = tf \times idf$). We choose to give higher weight to the idf value (i.e., $w = tf \times idf^2$). The reason behind this modification is to normalize the inherent bias of the tf measure in short documents. 

Finally, the set of feature vectors for the $N$ things in the dataset $\mathbf{\vec{v}} = [\vec{v}_1,...,\vec{v}_N]$ where  $\vec{v}_i \in \mathbb{R}^{m}$ is the feature vector for each thing, 
$m$ is the size of vocabulary we produced. 
For better performance, we perform a cosine normalization for tf-idf vectors: $\hat{\mathbf{v}} = \dfrac{\vec{\mathbf{v}}}{||\vec{\mathbf{v}}||_2}$ \cite{salton1988term}.

%
%
%


%

\paragraph{Building a discriminative classifier}
After obtaining the features based on attributes of $\mathbf{G}$ and things, we combine the features ($\mathbf{F_L} + \mathbf{F_S}$ + $\mathbf{F_C}$) together and feed them into a discriminative classifier. 

Our method is a very flexible feature-based method, where the structural features can be put into any discriminative classifier for classification. In this paper, we evaluate our method on SVM and Logistic regression. 
%
Specifically, we adopt LibSVM~\cite{chang2011libsvm} for one-vs-rest classification.


\paragraph{Discussion}
The high and real-time streams of interactions between human and ubiquitous things call for online processing techniques that are suitable to large scale datasets and 
can 
rapidly update
to reflect constantly evolving 
contextual similarities due to changing 
conditions of things (e.g., social networks, status, locations etc).
Our proposed model can be easily extended 
to deal with large scale IoT data streams with online-processing and incremental techniques due to the following characteristics of the model: 

\begin{itemize}
\item  We characterize each thing as a discriminative feature descriptor including static features (e.g., content-based features) and easily integrated dynamic features (e.g., locations, and instantaneous status of things). The feature vectors can be continuously updated with users' interactions over things in a real-time manner. 
It should be noted that we focus on training an offline model with mixture of static and dynamic features, which 
is possible to be leveraged for online learning. 
For instance, the model can be integrated 
in an incremental 
learning framework that is able to continuously update and learn from newly observed
data.

\item Our proposed model does not require any explicit input from users. All the contextual information is automatically obtained from users' social networks, localization techniques and sensor technologies in a non-obtrusive way. 

\item The process of contextual similarity calculation  works based on the random walk techniques, which has been successfully used in large-scale online search engines. This type of techniques can be easily parallelized (e.g., using Hadoop framework for improving performance) and processed in real time. 
\end{itemize}


\section{Evaluation}
\label{sec:evaluation}
In this section, we firstly describe our experimental settings, and then showcase the applicability and performance of our proposed technology based on 
feature-based things annotation. We also report the experimental results. 

\subsection{Data Acquisition}
\label{sec:data}
\begin{figure*}[!tb]
\begin{center}
\includegraphics[width=13cm]{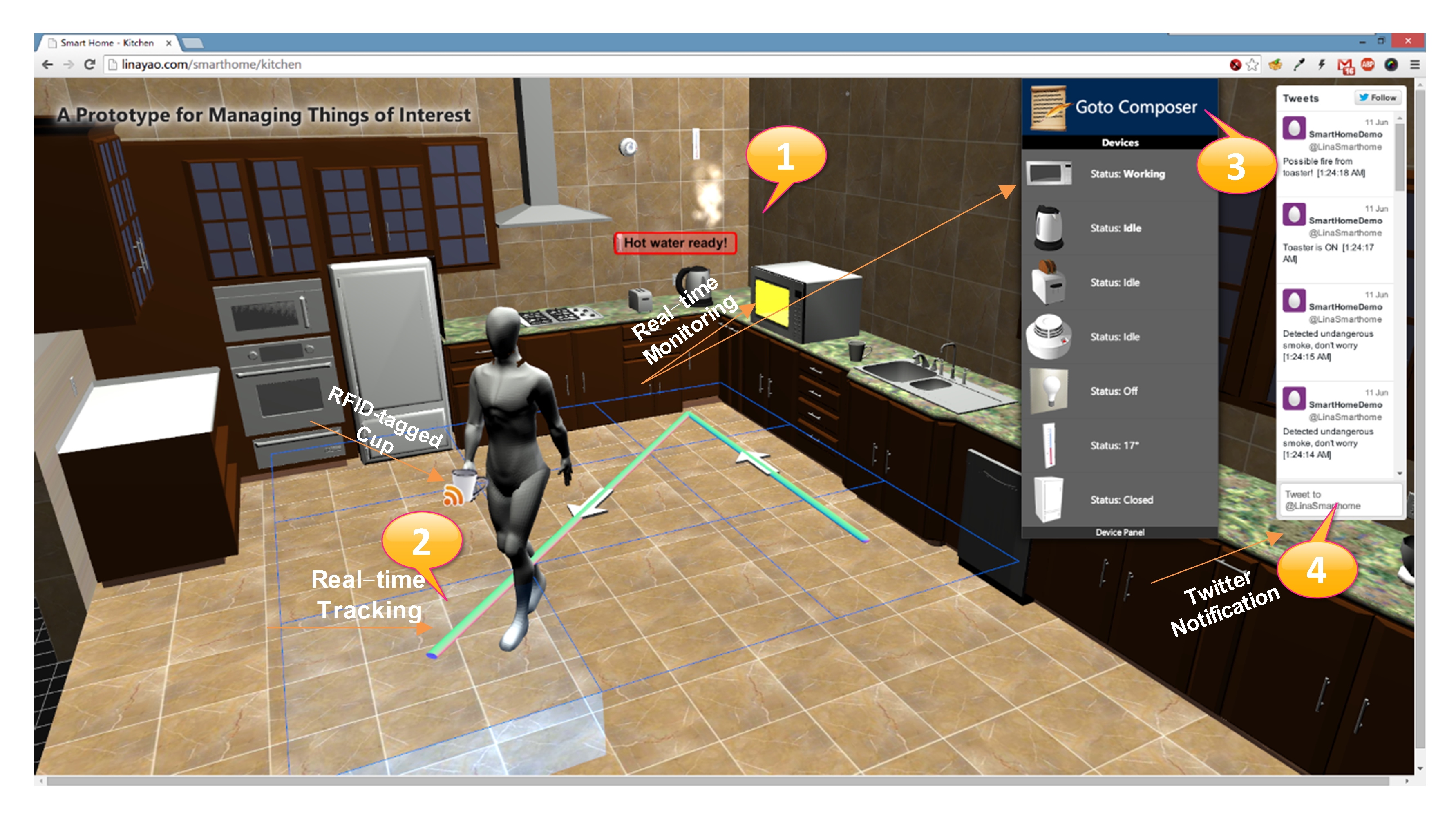}
\caption{Web-based prototype interface}
\label{fig:interface}
\end{center}
\end{figure*}

\begin{figure}[!h]
\begin{center}
\begin{minipage}{6cm}
\includegraphics[width=6cm]{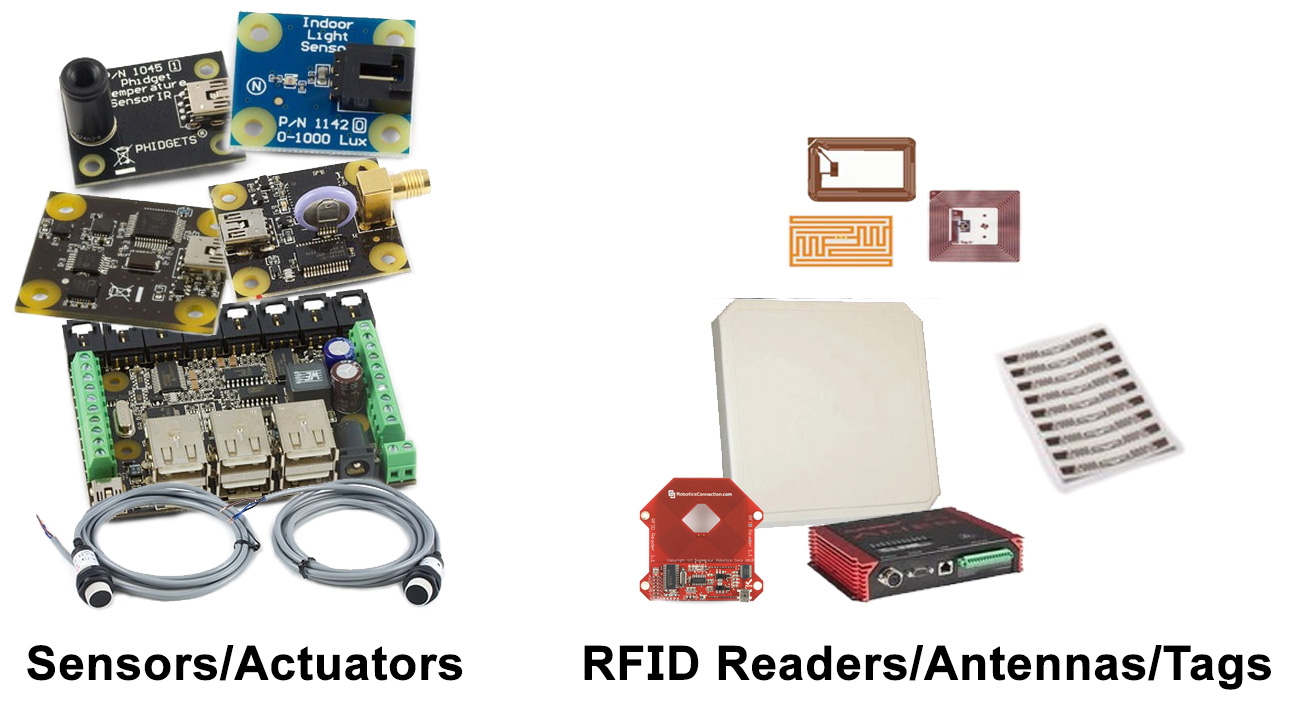}
\centering{(a)}
\end{minipage}
\hspace{5mm}
\begin{minipage}{6cm}
\includegraphics[width=6cm]{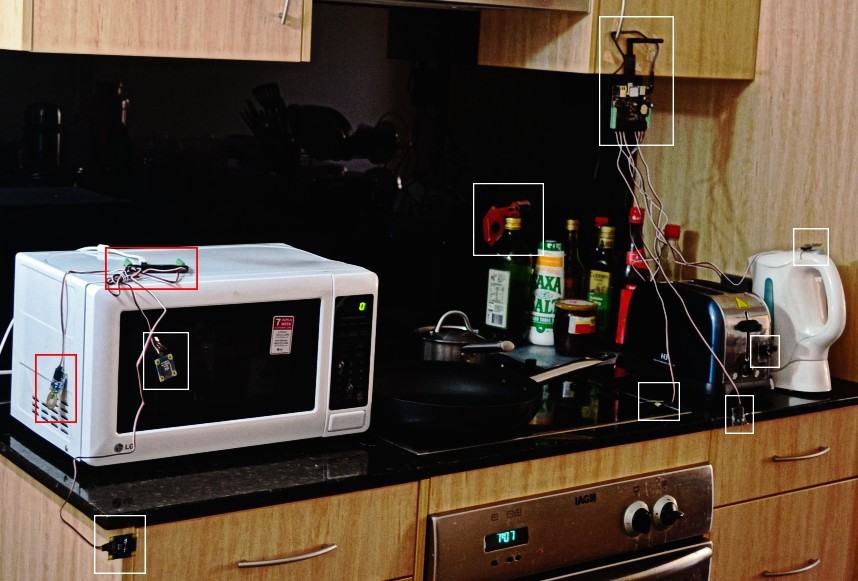}
\centering{(b)}
\end{minipage}
\caption{(a) some sensors and RFID devices; (b) WoT-enabled microwave oven}
\label{fig:device}
\end{center}
\end{figure}

\begin{figure}[!tb]
\begin{center}
\includegraphics[width=12cm]{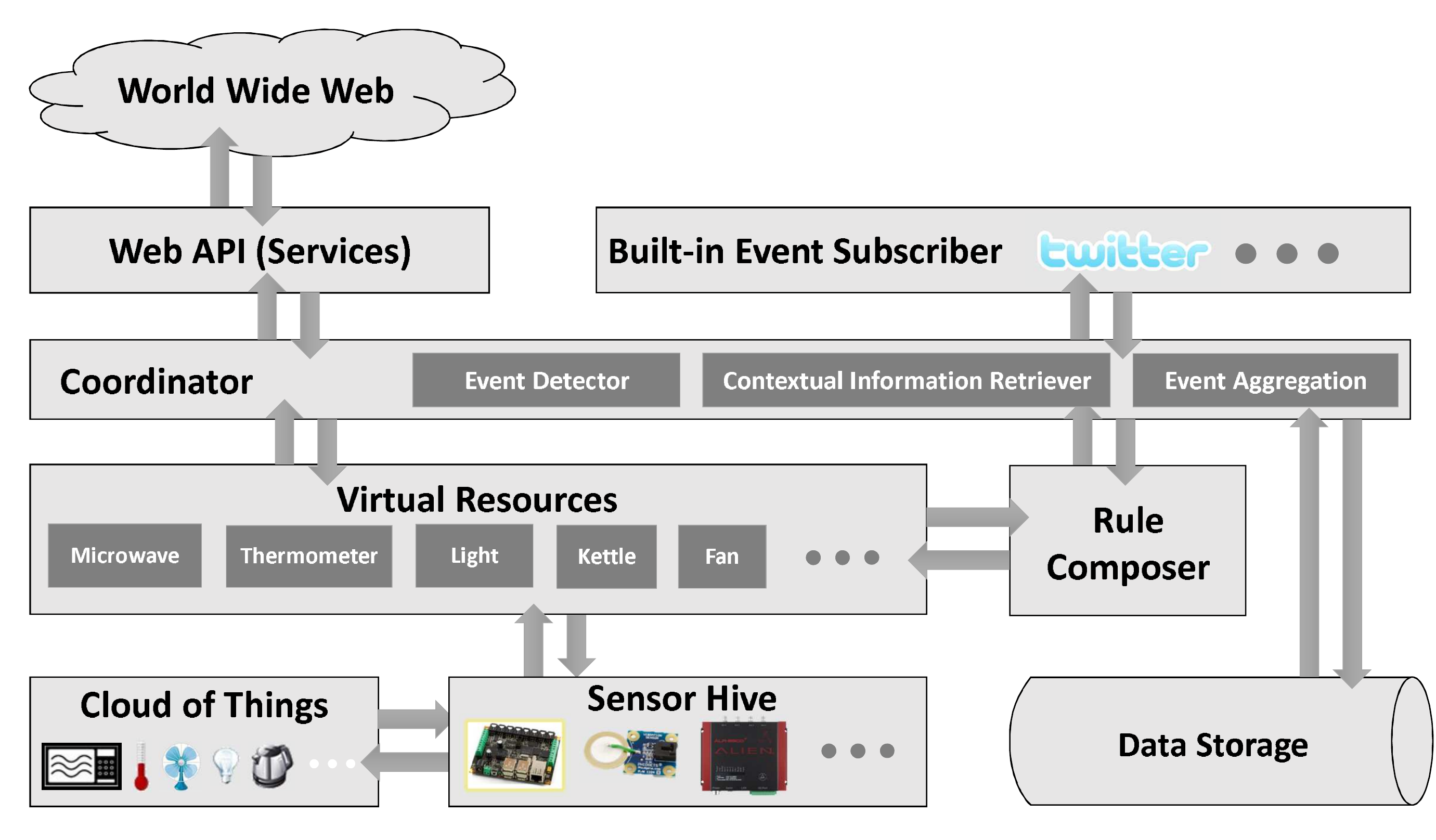}
\caption{System architecture of the testbed}
\label{fig:arch}
\end{center}
\end{figure}

%
Due to the lack of experimental public data sets, we set up a testbed that consists of several different places in the first author's home (e.g., bedroom, bathroom, garage, kitchen etc), where approximate 127 physical things (e.g., couch, laptop, microwave, fridge etc.) are monitored by attaching RFID and sensors. Table~\ref{tab:part1data} presents the statistics of things used in this paper. 
This task greatly benefits from our extensive experience in a 
large RFID research project~\cite{Wu-EDBT2012,Wu-tpds2013}. 
Figure~\ref{fig:interface} shows a research prototype we developed that provides an environment where 
users can check and control things in real time via a Web interface\footnote{https://www.youtube.com/watch?v=q5dZDZ3PZ9Y}.
Figure~\ref{fig:device} (a) shows some RFID devices and sensors used in the implementation and Figure~\ref{fig:device} (b) shows part of the kitchen setting in our testbed. 
In our implementation, things are exposed on the Web using RESTful Web services, which can be discovered and accessed from a Web-based interface. 
Figure~\ref{fig:arch} shows the architecture of our testbed. 

\begin{table}[!h]
\centering
\tbl{Dataset}{
\begin{tabular}{c l l l} \hline
No. & Category & \# Things & \# Labels\\ \hline\hline
1 & Entertainment& 28 & 118 \\ 
2 & Office & 20 & 51 \\ 
3 & Cooking & 25 & 103 \\ 
4 & Transportation & 11 & 24 \\ 
5 & Medicine/Medical & 10 & 18 \\ 
6 & Home Appliances & 33 & 83\\ \hline
\end{tabular}}
\label{tab:part1data}
\end{table}

To collect the records of things usage events, we need to figure out i) how to detect a usage event when it is happening; and ii) how to retrieve this thing's corresponding three contextual information. 

There are two ways to detect usage events of things with two identification technologies used, namely
{\em senor-based state changes} and {\em RFID-based mobility detection}.

\paragraph{Sensor-based state changes} The usage of a thing instrumented with sensors is reflected by the changes of the thing's status.
When the status is changed, the corresponding thing is used. For example, when the status of a microwave oven is turned from idle to working, we see that this oven is being used. 
For such event detections, we adopt sensors to track the state changes of things.

\paragraph{RFID-based mobility detection} 
We determine whether the RFID-enabled things are in use via detecting their mobility. 
The movement of a thing indicates that the thing is being used. For example, if a coffee mug is moving, it is likely that the mug is being used. 
For such detections, we adopt a generic method based on comparing descriptive statistics of the Received Signal Strength Indication (RSSI) values from RFID readers in consecutive sliding windows \cite{parlak2011activity}. The statistics obtained from two consecutive windows are expected to differ significantly when a thing is mobile. A threshold can be set to determine whether this difference is related to a mobility and can be regarded as a valid usage event.

Each usage event is associated with identity (user), temporal (timestamp) and spatial (location) information. To obtain the user information, 
in our current work, 
we use a manual labeling method where each participant needs to mark and record their activities. 
For the temporal information, 
we choose to 
divide the time of one day into 24 equal intervals. Each interval is one hour. 
If the timestamps of a usage event collected is 9:07am, it will be assigned into the temporal cluster between 9:00am to 10:00am. It should be noted that other equal intervals (e.g., half hour for an interval) are also applicable to our approach. 

To get the localization information, which indicates where a thing is when it is used. In the localization step, our aim is to identify the coarse-grain locations, the zone where the object lies. 
We need to consider two situations for things, {\em static} and {\em mobile}. For static things (e.g., refrigerator, microwave oven), the location information of such things is prior knowledge. For mobile things (e.g., RFID-tagged coffee mug), we provide coarse-grain or fine-grain location information. For the coarse-grain method, since the Received Signal Strength Indication (RSSI) signal received from a tagged thing reveals its proximity to an RFID reader antenna. We divide an area into multiple zones and each zone is covered with a mutually exclusive set of RFID antennas. The zone scanned by the antenna with the maximum RSSI is taken to be the thing's location. For the fine-grain method, it is determined by comparing the signal descriptors from a thing at unknown location to a previously constructed radio map or fingerprints. We use the Weighted $k$ Nearest Neighbors algorithm (w-kNN), where we find the most similar fingerprints and compute a weighted average of their 2D positions to estimate the unknown tag location~\cite{ni2004landmarc}. 



To conduct experimental studies, we manually labeled 127 things with 397 different labels. It should be noted that some things belong to multiple categories, therefore having multiple labels. 
For example, a $\textit{Wii}$ device belongs to category label $\textit{Entertainment}$ as well as $\textit{Home Appliance}$. 
This dataset serves as the ground-truth dataset in our experiments for performance evaluation.
Ten volunteers participated in the data collection phase by interacting with RFID tagged things for a period of four months, generating 20,179 records on the interactions of the things tagged in the experiments.

\subsection{Metrics} 
We use micro-F1 and macro-F1 as evaluation measures. The F1 measure is the harmonic mean of $Precision$ (P) and $Recall$ (R), which can be calculated as: $F_1 = 2\dfrac{P \times R}{P + R}$. The Micro-F1 is defined as:
 
\begin{equation}
Micro-F1 = \dfrac{2\sum_{j=1}^{c}\sum_{i=1}^{n}\hat{y}_{i}^
{c}y_{i}^{c}}{\sum_{j=1}^{c}\sum_{i=1}^{n}\hat{y}^{c}_{i} +
\sum_{j=1}^{c}\sum_{i=1}^{n}y_{i}^c}
\end{equation}
 
where $n$ is the number of testing data, $c$ denotes the number of category labels, $y_i$ is the true label vector of the $i$-th sample, $y_{i}^j = 1$ if the instance
belongs to category $j$, $-1$ otherwise. $\hat{y_{i}}$ is the predicted label vector. The micro-F1 measure weights equally on all samples, thus favoring the performance on common category labels. Macro-F1 is calculated as mean arithmetical value for F1 on each label. It measures weights equally on all the category labels regardless of how many samples belong to it, thus favoring the performance on rare category labels. Macro-F1 is defined as: 

\begin{equation}
Macro-F1 = \dfrac{2\sum_{i=1}^{n}\hat{y}_{i}^
{c}y_{i}^{c}}{n^{2}|c|(\sum_{i=1}^{n}\hat{y}^{c}_{i} +
\sum_{i=1}^{n}y_{i}^c)}
\end{equation}


\subsection{Experimental Results}
\label{sec:experiment}
In this section, we study the performance of our proposed DisCor-T approach based on things annotation described in Section~\ref{sec:applications}. In particular, we will report the evaluation results for things annotations in terms of i) sensitivity analysis on varying weight value $\alpha$ and $\beta$ in Equation~\ref{equ:overall}; ii) overall performance with different configurations of features, and iii) impact on introducing spatio-temporal integrity in our approach. 



\subsubsection{Parameters Tuning}
\begin{figure}[!tb]
\begin{center}
\begin{minipage}{4.5cm}
\includegraphics[width=4.5cm]{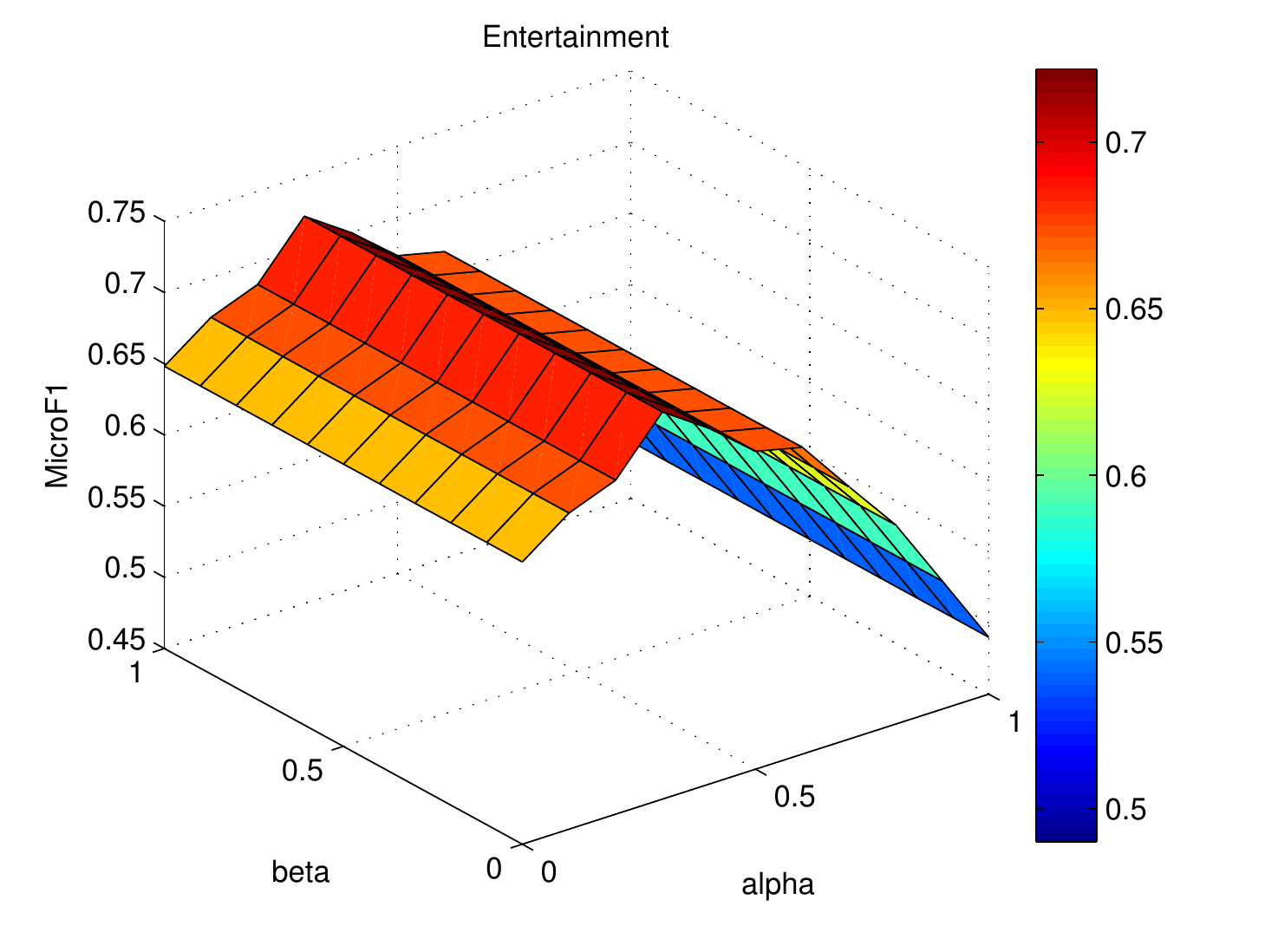}
\centering{}
\end{minipage}
\begin{minipage}{4.5cm}
\includegraphics[width=4.5cm]{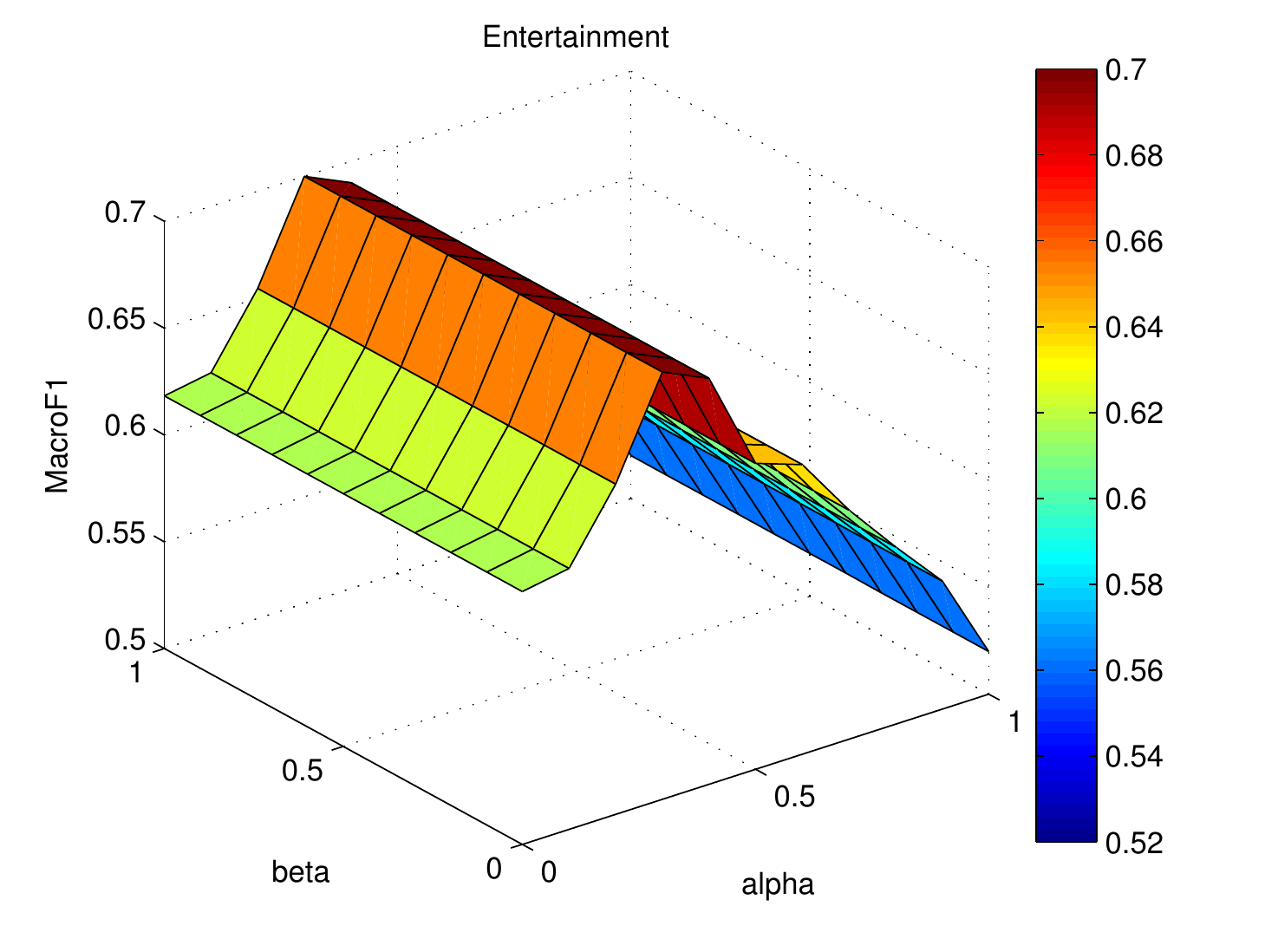}
\centering{}
\end{minipage}
\begin{minipage}{4.5cm}
\includegraphics[width=4.5cm]{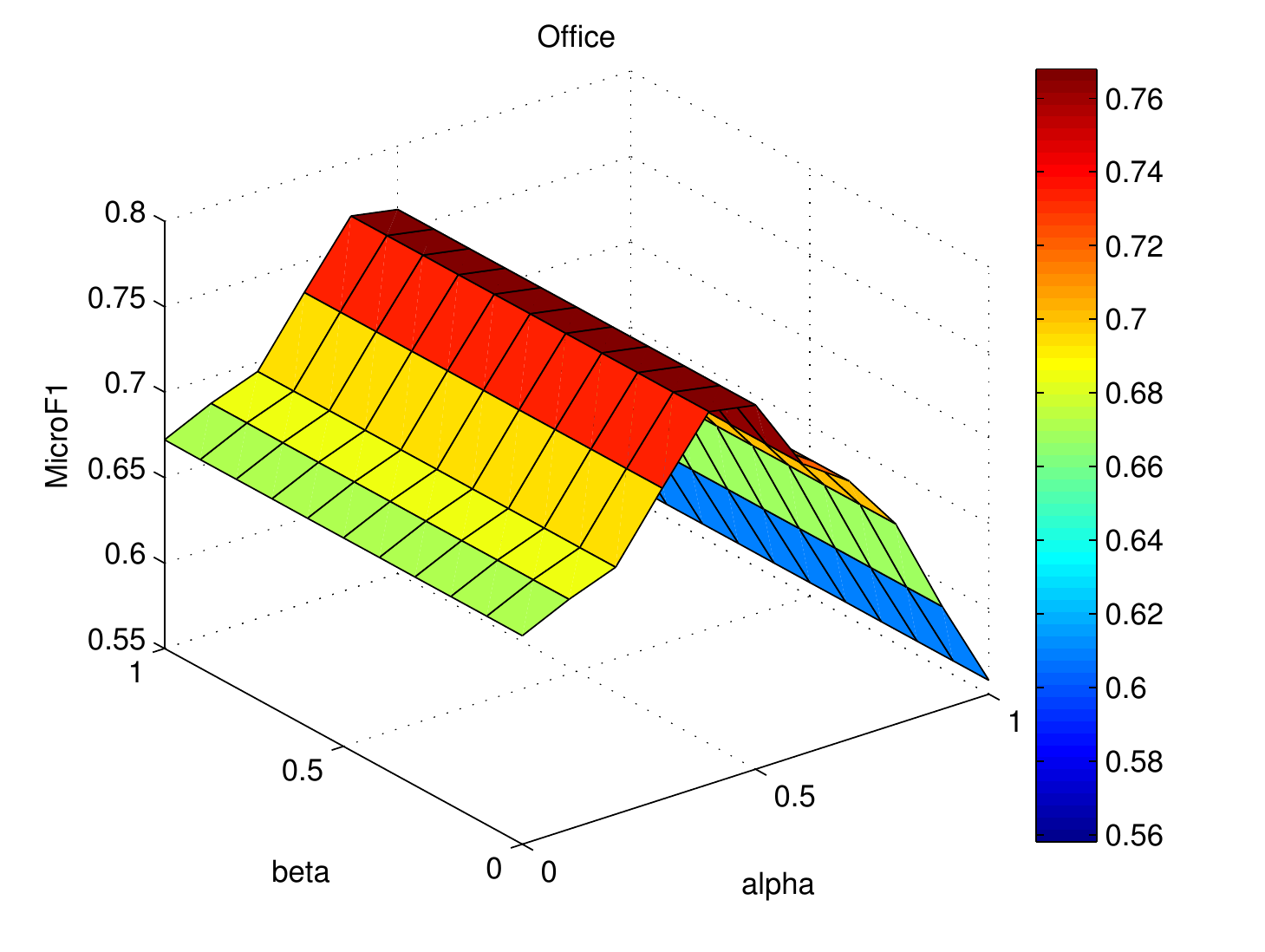}
\centering{}
\end{minipage}
\begin{minipage}{4.5cm}
\includegraphics[width=4.5cm]{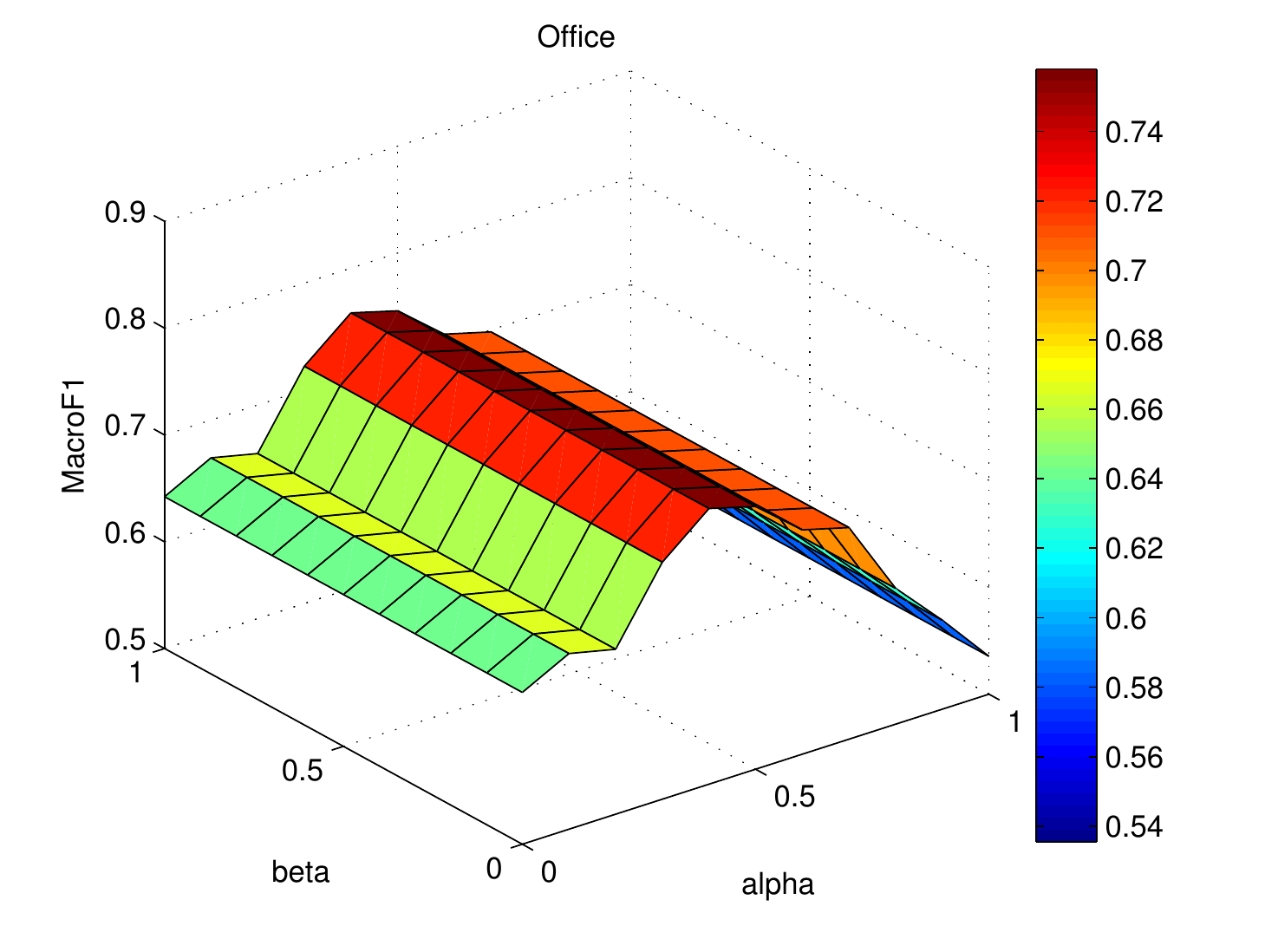}
\centering{}
\end{minipage}
\begin{minipage}{4.5cm}
\includegraphics[width=4.5cm]{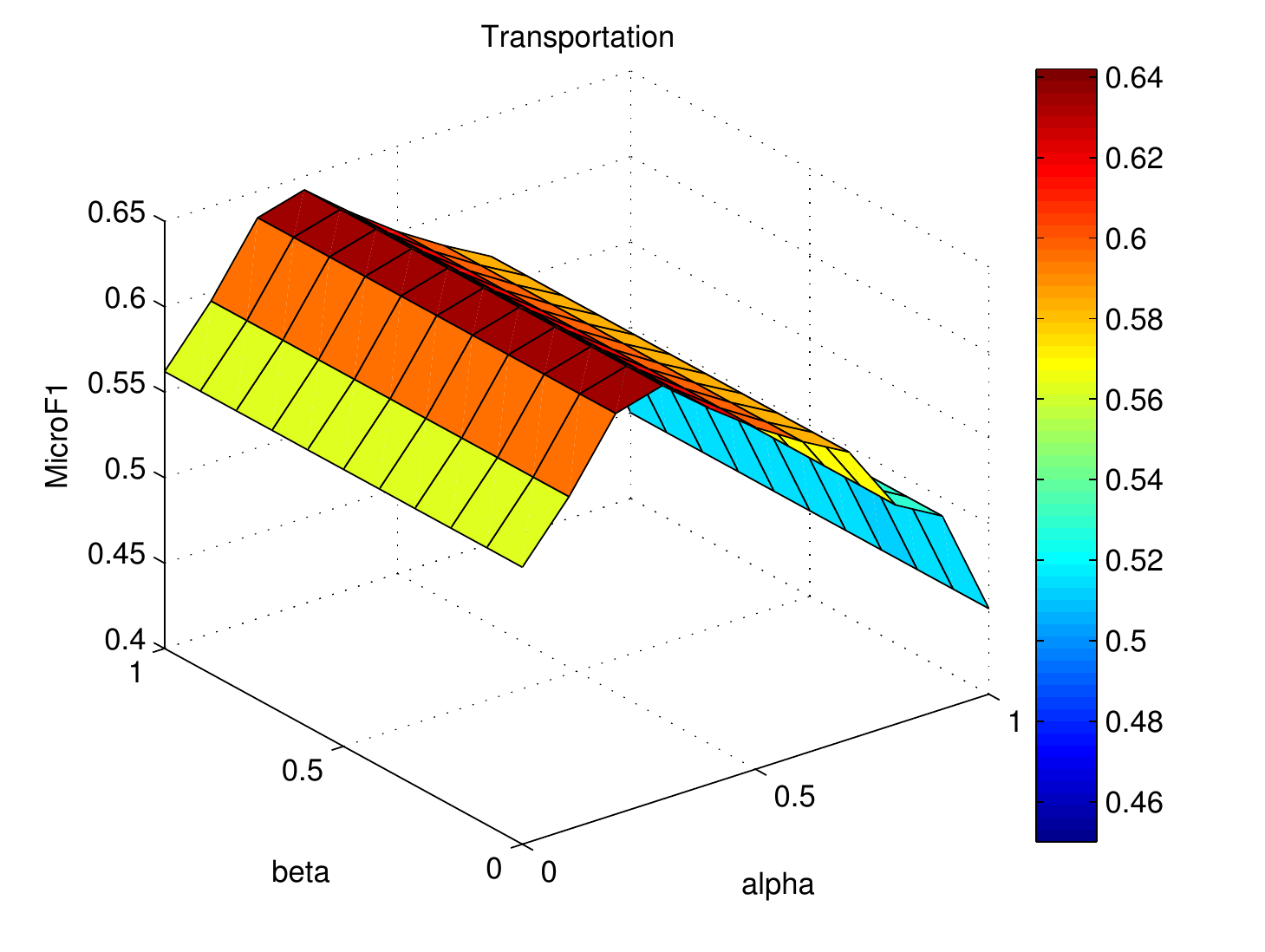}
\centering{}
\end{minipage}
\begin{minipage}{4.5cm}
\includegraphics[width=4.5cm]{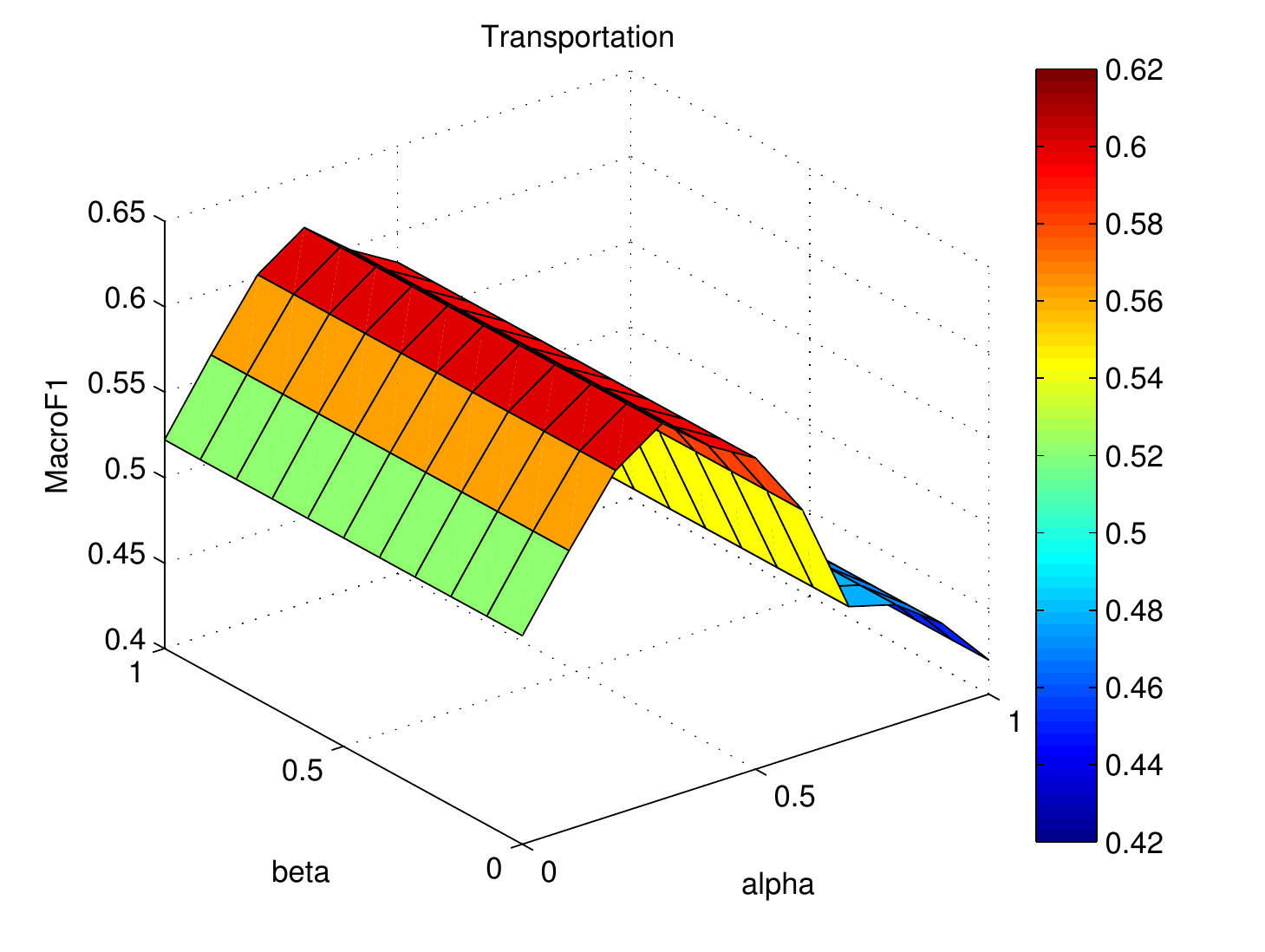}
\centering{}
\end{minipage}
\begin{minipage}{4.5cm}
\includegraphics[width=4.5cm]{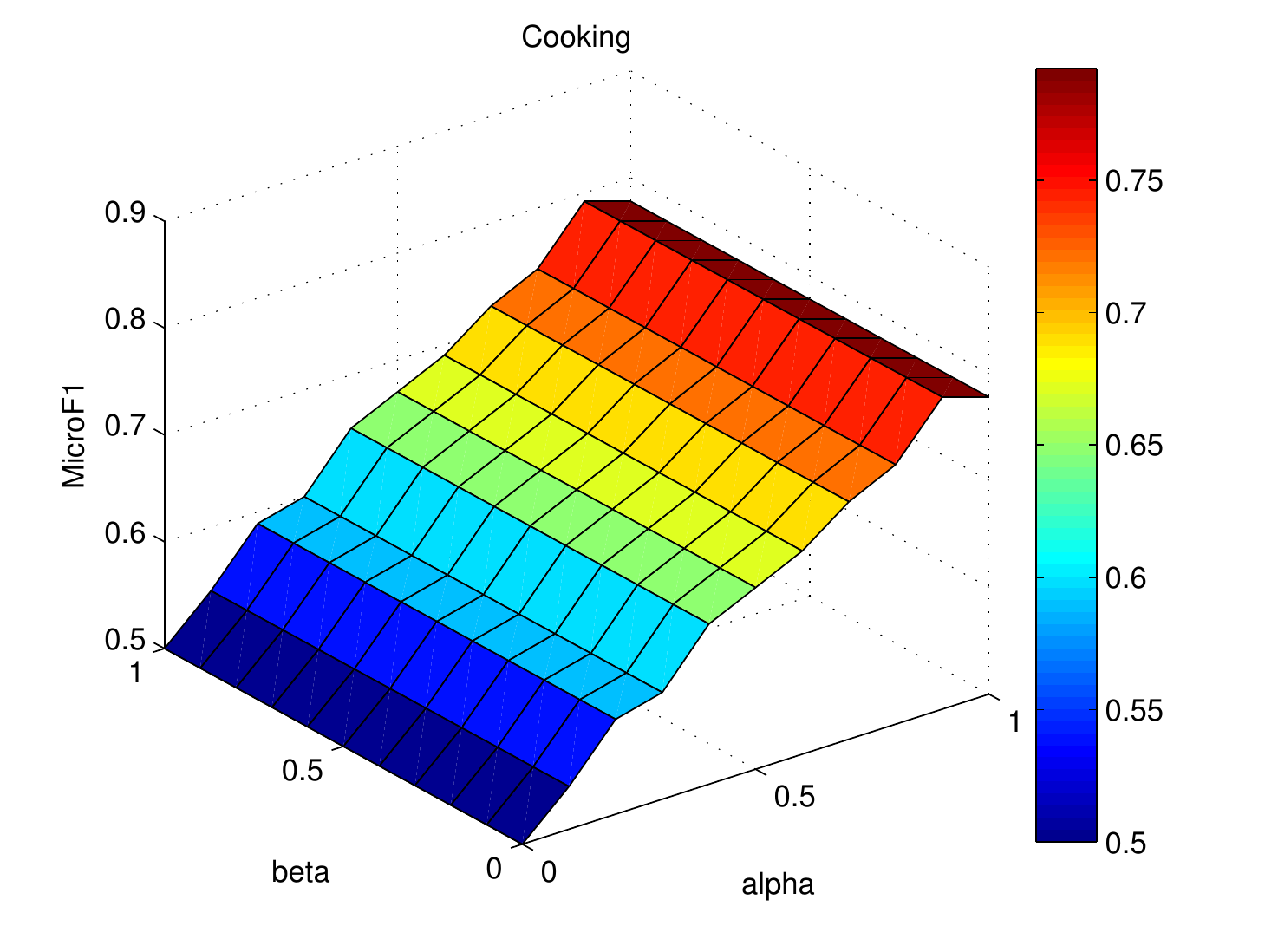}
\centering{}
\end{minipage}
\begin{minipage}{4.5cm}
\includegraphics[width=4.5cm]{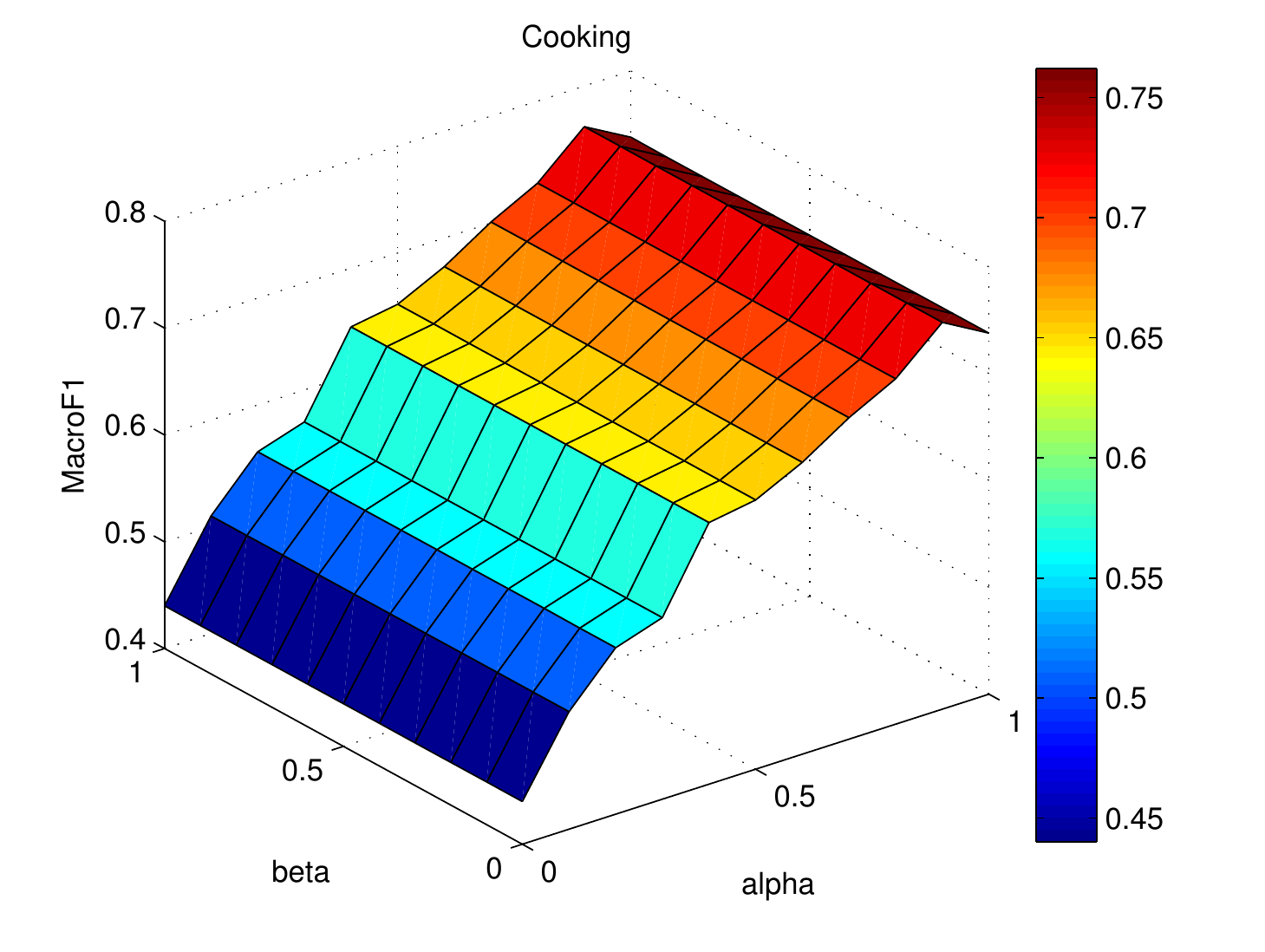}
\centering{}
\end{minipage}
\begin{minipage}{4.5cm}
\includegraphics[width=4.5cm]{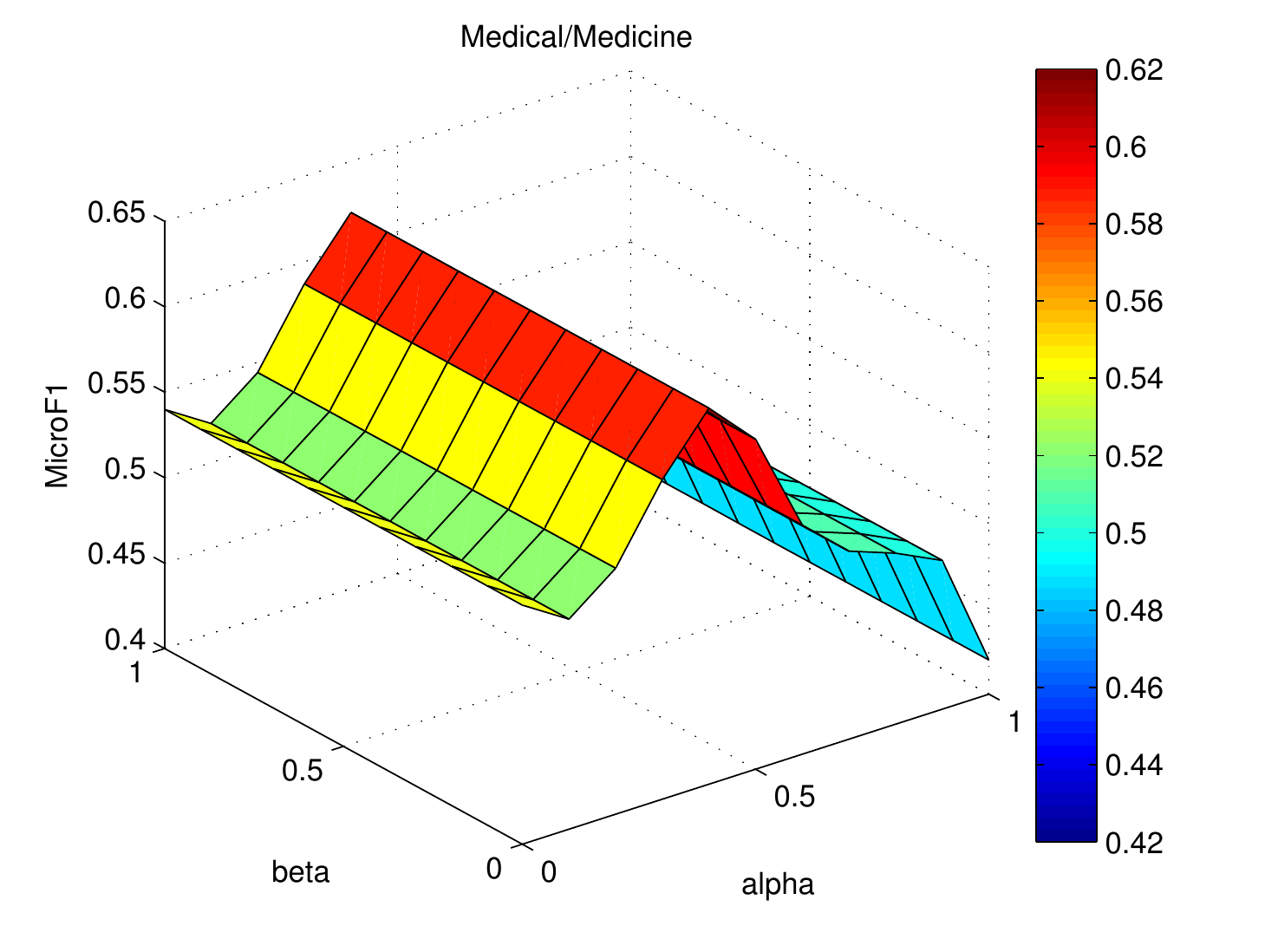}
\centering{}
\end{minipage}
\begin{minipage}{4.5cm}
\includegraphics[width=4.5cm]{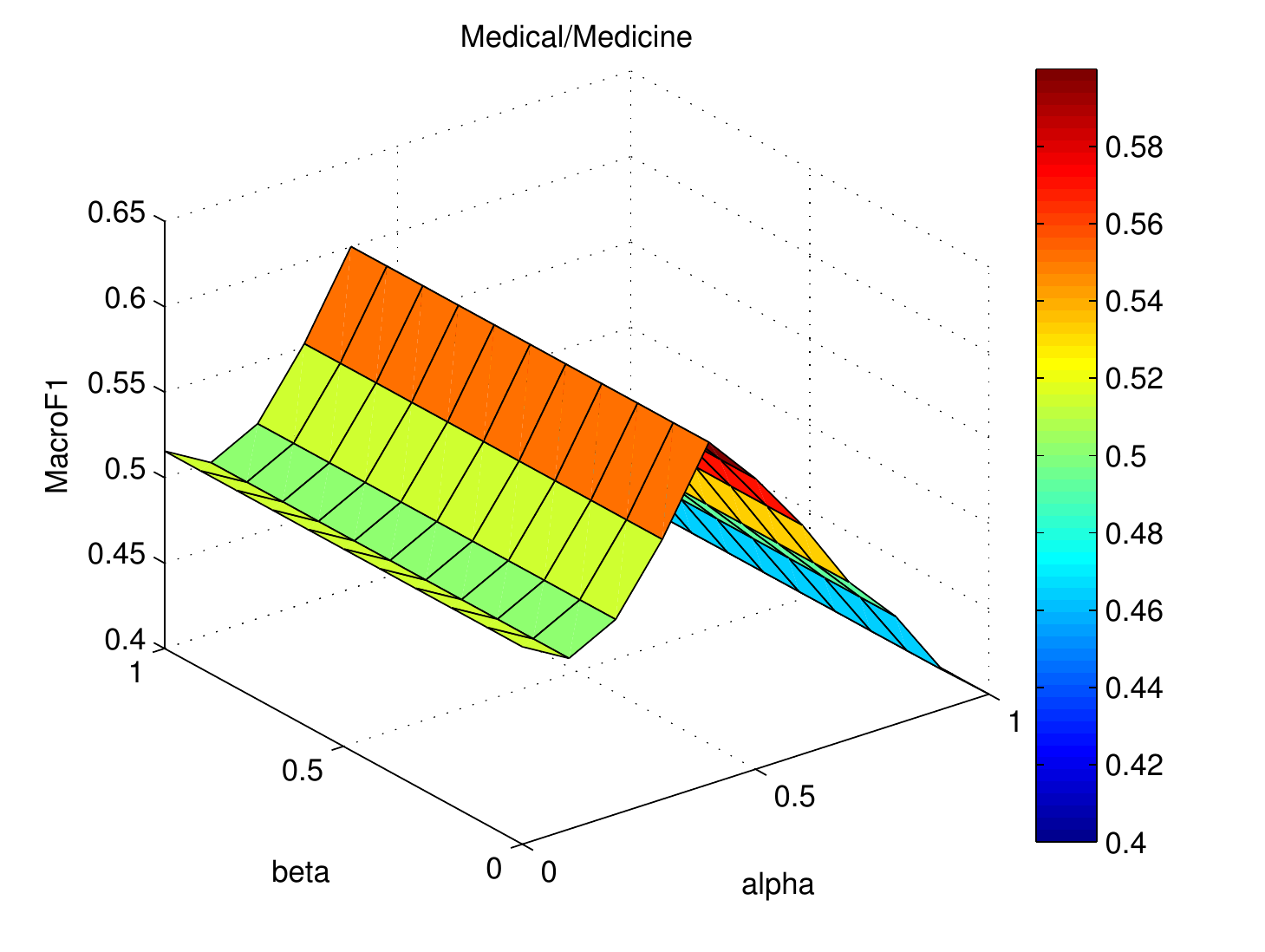}
\centering{}
\end{minipage}
\begin{minipage}{4.5cm}
\includegraphics[width=4.5cm]{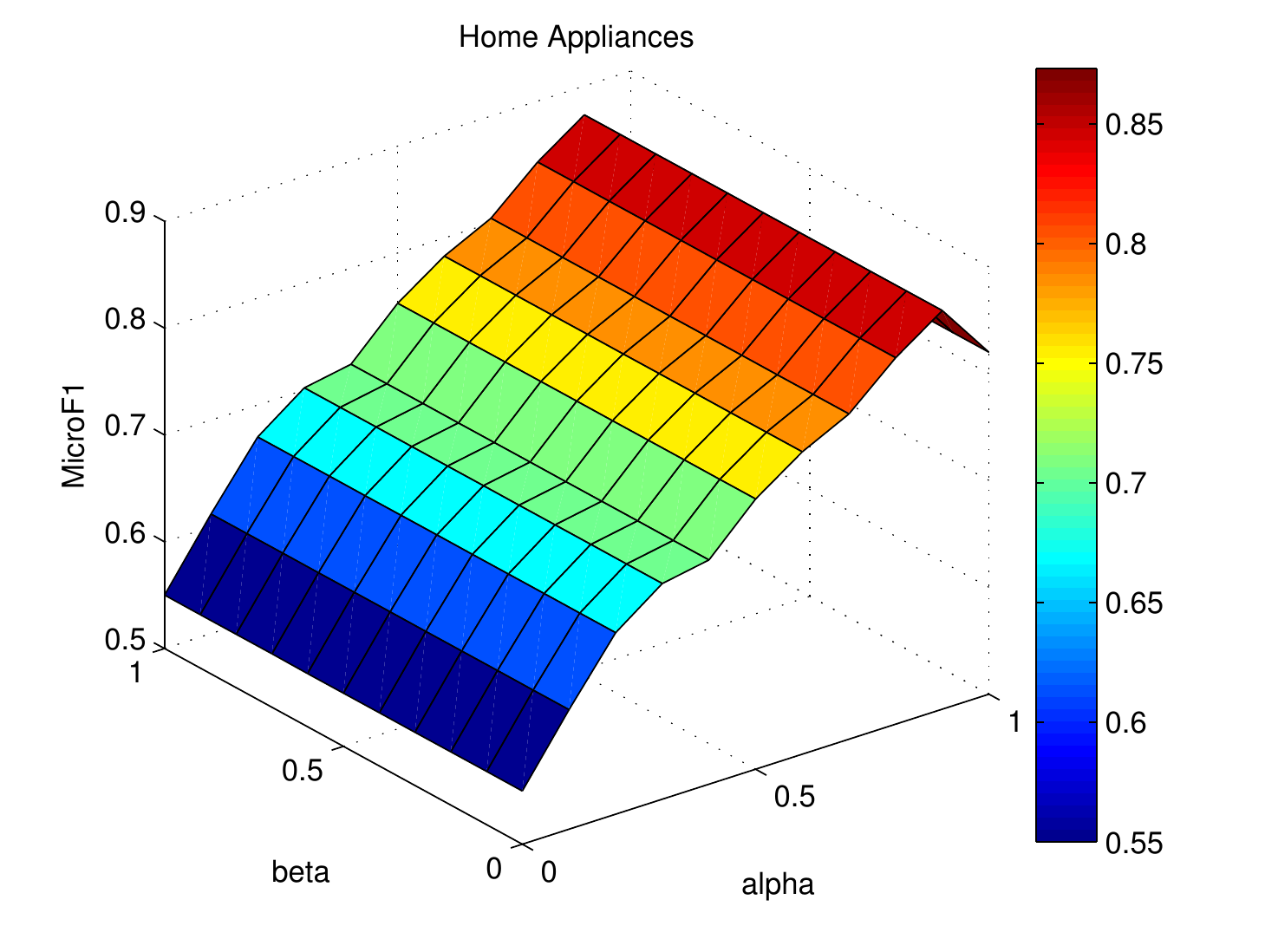}
\centering{}
\end{minipage}
\begin{minipage}{4.5cm}
\includegraphics[width=4.5cm]{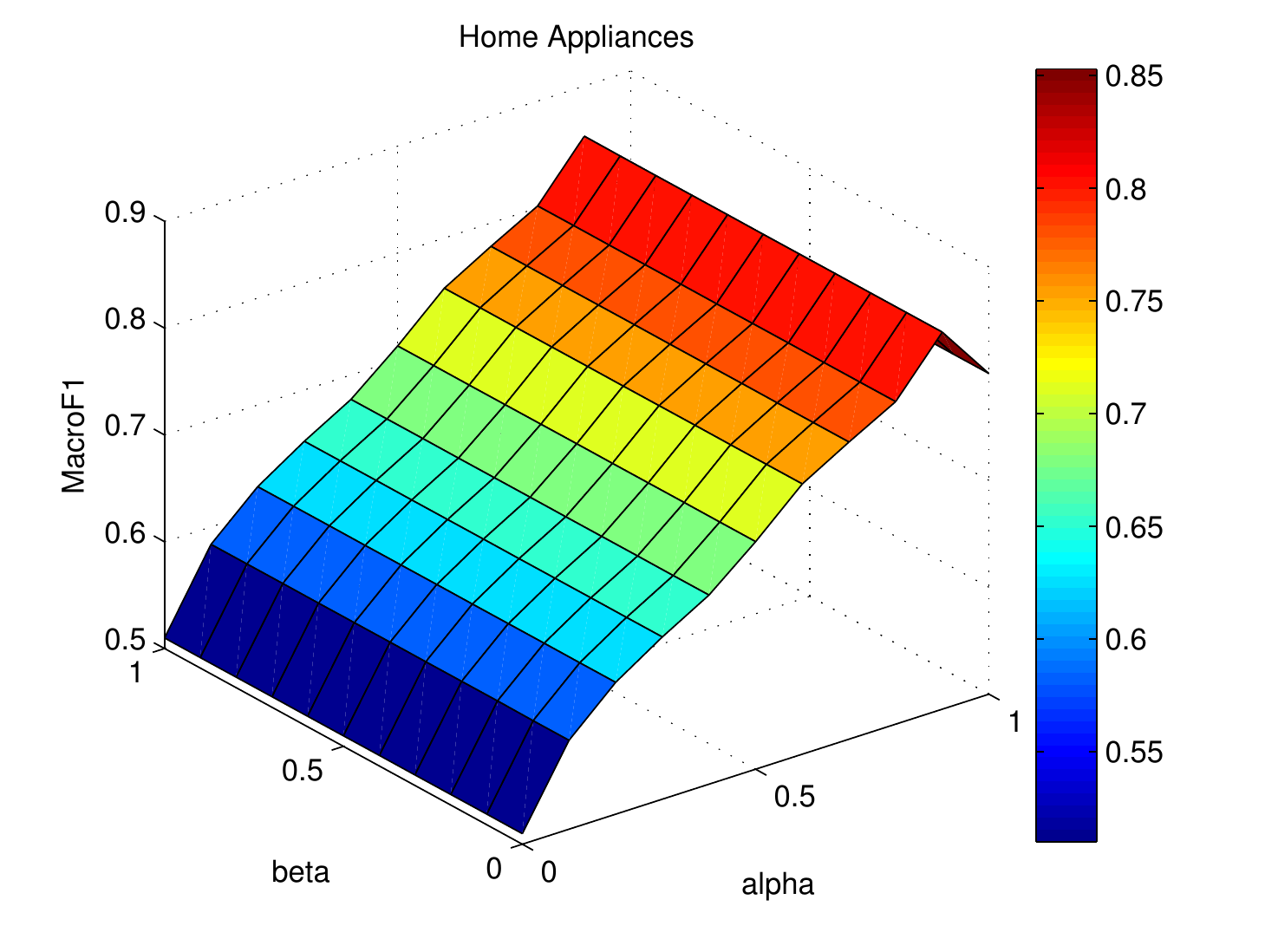}
\centering{}
\end{minipage}

\caption{Parameters tuning measured by MicroF1 and MacroF1 with different $\alpha$ and $\beta$ on six categories}
\label{fig:tuning}
\end{center}
\end{figure}

This experiment 
aims at studying the impact of 
tuning parameters $\alpha$ and $\beta$ in Equation~\ref{equ:overall} on different categories of things. We varied the $\alpha$ from $0.1$ to $0.9$ increment with $0.1$ each time, while $\beta$ was varied from 0.9 to 0.1 decrement with $0.1$ each time, and implemented our annotation algorithm on the produced graph to evaluate the annotation performance. The results on the six categories of things are shown in Figure~\ref{fig:tuning}. 

We can see some interesting patterns from the figures. For instance, with bigger temporal-spatial weight $\alpha$ and smaller social weight $\beta$, the annotation algorithm has 
better performance on \texttt{Cooking} and \texttt{Home Appliance} categories. It means both of these categories are sensitive to the temporal-spatial information but not 
the user aspect, i.e., the impact of $\beta$ on classification of these categories is very limited. 
The possible reason is that things in these categories are connected by tight contextual relevance for their regular users. As a result, there presents little improvement 
when increasing the weight of the user aspect. On the contrary, we observe that for categories \texttt{Entertainment} and \texttt{Transportation}, the user aspect shows obvious impact since better performance is obtained when $\beta$ increases. The possible reason is that these categories show some obvious convergence for common users. For the categories of \texttt{Office} and \texttt{Medicine/Medical}, they do not possess obvious preference over social or contextual (temporal-spatial) information. For example, it is hard to find a common time for people to receive initial treatment of injuries or illnesses at work place, which usually happen randomly.  
We also can observe that the performance is not sensitive to varying $\beta$ across all the categories. The reason might lie in the number of users is not big enough to carry discriminative information to differentiate the impact of social graph. Conducting more experiments using large-scale real world WoT data will be one of our future works. 


\begin{table}[!h]
\centering
\tbl{$\alpha$ and $\beta$ configuration}{
\begin{tabular}{l l l} \hline
Category & $\alpha$ & $\beta$\\ \hline\hline
Entertainment& 0.4 & 0.6 \\ 
Office & 0.5 & 0.5 \\ 
Cooking & 0.8 & 0.2 \\ 
Transportation & 0.4 & 0.6 \\ 
Medicine/Medical & 0.5 & 0.5 \\ 
House Appliances & 0.9 & 0.1\\ \hline
\end{tabular}}
\label{tab:tuning}
\end{table}

%

\subsubsection{Overall Performance}

This experiment evaluates the performance of
things annotation described in Section~\ref{sec:rwr}. We randomly removed the category tags of a certain percentage, ranging from  10\% to 50\%, of things from each category of the ground-truth dataset. These things were used to test our approach while the rest were used as the training set. 
We iterated five times for each training percentage and took the averaged value as the final result. 
Our algorithm produces a vector of probabilities, representing the assignment probabilities of all labels for 
an unknown object. In our experiments, we ranked these probabilities and chose the top $k$ labels to compare with the ground truth labels. The $k$ value was set to the number of ground truth labels for each 
unknown object and it varies from object to object. The parameters $\alpha$ and $\beta$ were set as 0.5 each.

\begin{figure}[!tb]
\centering
\begin{minipage}{6.9cm}
\includegraphics[width=6.9cm]{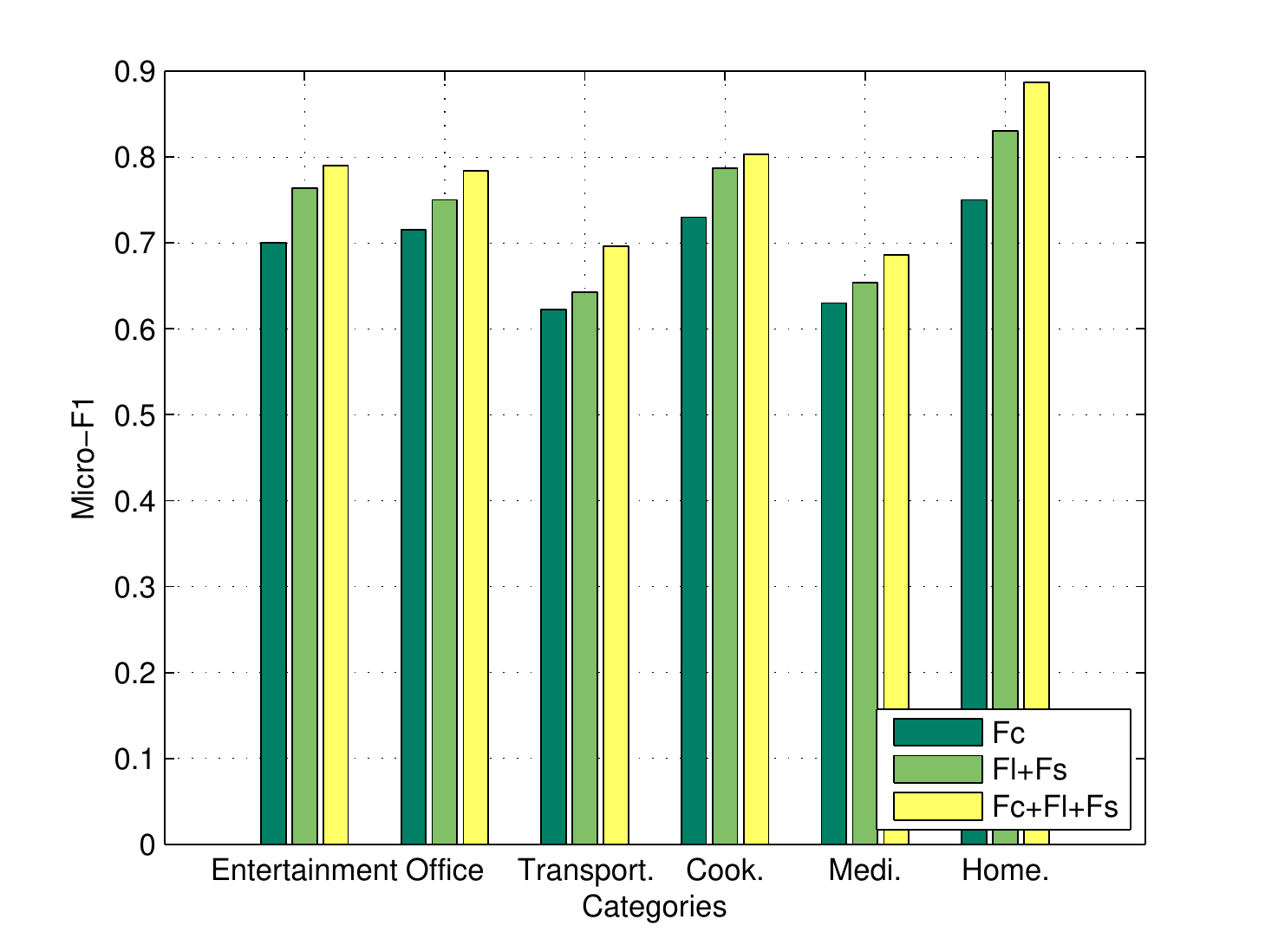}
\centering{(a)}
\end{minipage}
\begin{minipage}{6.9cm}
\includegraphics[width=6.9cm]{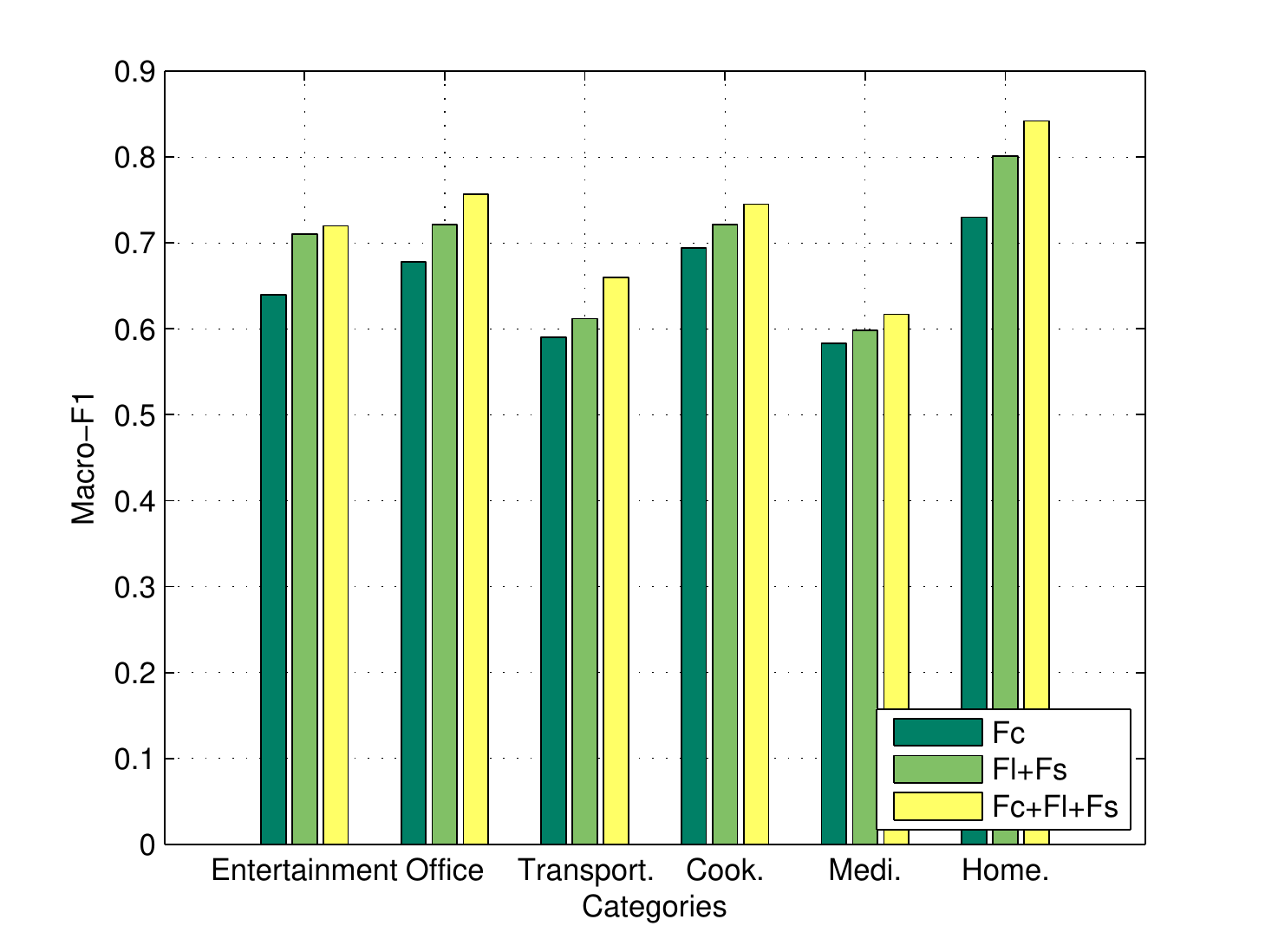}
\centering{(b)}
\end{minipage}
\caption{Overall performance comparison: (a) Micro-F1, (b) Macro-F1}
\label{fig:overall}
\end{figure}

We particularly compared the annotation performance by using i) the features obtained from $\mathbf{G}$, ii) the features obtained from thing descriptions (i.e., content features $\mathbf{F_C}$), and iii) the combination of the both. Each process was repeated 10 times and the average results were recorded. Similar observations were obtained for different testing percentages. Figure~\ref{fig:overall} shows the result when we removed 30\% of things from each category of the ground-truth dataset.

Descriptions of things are normally short and noisy, it is therefore not surprising that the performance based on content features only is worse than the one based on implicit structural features (i.e., $\mathbf{F_L} + \mathbf{F_S}$) in most categories. The consistent good performance from the latent features also indicates that our top-$k$ correlation graph $\mathbf{G}$ is able to capture the correlations among things well. 
From the figure, we can see that by combining the two together, the performance of all six categories is increased and is the best consistently among the three. 

\begin{table}[!tb]
\begin{center}
\begin{small}
\tbl{Performance comparison with STI and without STI}{
\begin{tabular}{l l l l l l l} \hline
\bf Category &\multicolumn{2}{c}{\bf Entertainment} &\multicolumn{2}{c}{\bf Office} &\multicolumn{2}{c}{\bf Transportation} \\ \hline \hline
 & Micro-F1 & Macro-F1 & Micro-F1 & Macro-F1 & Micro-F1 & Macro-F1\\ \hline
\bf Without STI &0.6954& 0.6533 &0.7442& 0.7201& 0.6333 & 0.6232 \\ 
\bf With STI &0.7226& 0.6818 &0.7854 & 0.7667& 0.6613& 0.6648\\ 
\hline
\bf Category  & \multicolumn{2}{c}{\bf Cooking} &\multicolumn{2}{c}{\bf Medical/Medicine} &\multicolumn{2}{c}{\bf Home Appliance}\\ \hline \hline
 & Micro-F1 & Macro-F1 & Micro-F1 & Macro-F1 & Micro-F1 & Macro-F1\\ \hline
\bf Without STI & 0.7634 & 0.7213 & 0.6121 & 0.5900 & 0.8351 & 0.8113 \\ 
\bf With STI &0.7987& 0.7451 & 0.6162& 0.6004 &0.8876 & 0.8589\\ \hline
\end{tabular}}
\label{tab:sti}
\end{small}
\end{center}
\end{table}

\subsubsection{Impact of Integrating Spatio-Temporal Information}

As indicated in Section~\ref{sec:context}, user interactions with physical things usually present strong spatial-temporal correlations. In our approach, we treat spatial and temporal information of things usage events inseparable 
and believe that this integration
would offer better performance in discovering correlations among things. 

To validate this idea, we constructed two independent graphs based on time and location information from things usage events. Then random walk with restart (RWR) was performed in these two graphs separately. Together with the constructed social graph, a relational graph of things was constructed as described in Section~\ref{sec:rwr}. We label this approach as {\em No-STI} (without spatio-temporal integration) and our approach as {\em STI}.
In this experiment, 
we focus on studying the impact of the setting of $\alpha=1$ and $\beta=0$, which indicates that we only derive 
$\mathbf{R}$ based on 
the spatial-temporal graph (see Equation~\ref{equ:overall}), and we also compare it with our previous work \cite{lina2012cikm}, where we constructed two independent graphs based on time and location information, and then sum them up to get the overall relevance. In this way, 
the spatio-temporal information are treated independently.


We performed things annotation by using features obtained from two different relational graphs of things and
Table~\ref{tab:sti} shows the results when we removed 30\% of things from each category of the ground-truth dataset. 
The table clearly shows that the annotation performance is enhanced for almost all categories by introducing spatio-temporal integrity and  
\texttt{Medical/Medicine} is the only exception.
The reason is that user interactions with things in this category 
do not have strong connections with spatio-temporal patterns. In other words, people usually do not show periodic patterns when accessing medical related things (e.g., only when they are sick).   

\section{Related Work}
\label{sec:relatedwork}
In this section, we review some existing research efforts that are closely related to our work.

\subsection{Relational Learning}
Relational learning refers to the classification in a context where things or entities present multiple relations~\cite{tang2009relational}. One main technique on relational learning is based on the Markov assumption, where the labels of a node in a relational network are determined by the labels of nodes in its neighborhood. Collective inference \cite{angelova2006graph,jensen2004collective} and semi-supervised learning on graphs \cite{zhu2003semi} work on this assumption, which is constructed based on the relational features of labeled data, followed by an iterative process (e.g., relaxation labeling method) to determine class labels for unlabeled data. In \cite{ye2011semantic}, Ye et al. applied this methodology in location-based social  networks for deriving label probabilities for places.  The authors used the collective classification method that learns labels from the neighborhood, which only includes the nodes that hold the top-$k$ relevance with the prediction node. Collective inference and semi-supervised learning on graphs are limited in capturing local dependencies of nodes in the relational network. 

Some improvement on semi-supervised learning algorithms focused on the dependency between labels \cite{liu2006semi}, 
while some other work tried to capture the long-distance relevance of nodes. For example,~\cite{miller2009nonparametric} proposed a nonparametric latent feature models for link prediction.In~\cite{neville2005leveraging}, Neville and Jensen used clustering algorithm to find cluster membership and fix the latent group variables for inference.  

There are several works aiming to explore the relations in a heterogeneous network. For instance, Kong et al. \cite{kong2012meta} proposed a meta-path based collective classification approach, which exploits the multi-type dependencies of linked objects via interconnecting with different linkage paths. Sun et al. \cite{sun2012will} developed a meta-path based approach for relation prediction by considering target relation and topological information.
%
These approaches are either not suitable for networks, such as Web of Things (WoT) that contains a large number of things, where computational costs for inference are prohibitive, or not fully taking rich contextual information into account.

In our work, we extend the model to the relational network of things where a thing's usage history not only indicates user and temporal information, but also location information. As a result, a better performance in deriving latent features from the relational network of things can be achieved. In particular, we explore the relation between spatial information and temporal information by exploring the periodical pattern in human interactions on things. 

%
%

\subsection{Ubiquitous Things Searching}

Finding related and similar things is a key service and the most straightforward method of finding related things is the traditional keyword-based search, where user querying keyword is matched with the extracted description of things including textual descriptions on thing's functionalities and non-functional properties. 
For example, in Microsearch \cite{tan2010microsearch} and Snoogle \cite{wang2008snoogle}, each sensor is attached to a connected object, which carries a keyword-based description of each object. Following an ad hoc query consisting of a list of keywords, the system returns a ranked list of the top $k$ entities matching this query. As we pointed out, this method can not work well for ubiquitous things due to unique characteristics, e.g., insufficient description of things, inconsistency of the meaning of the textual information, more importantly this solution does not make use of implicit inter-correlations between things and their rich contextual information. 

Another mainstream solution is via semantic Web related techniques. Such solutions
typically use the meta data annotation (e.g., details related to a sensor such as sensor type, manufacturer, capability and contextual information), then use a query language to search related available things \cite{mietz2013semantic,christophe2011searching}.
Online sensors such as Pachube\footnote{https://pachube.com/}, GSN \cite{aberer2006middleware}, Microsoft SensorMap \cite{nath2007sensormap} and linked sensor middleware \cite{le2011linked} support search for sensors based on textual metadata that describes the sensors (e.g., type and location of a sensor, functional and non-functional attributes, object to which the sensor is attached), which is manually entered by the person who deploys the sensor. Other users can then search for sensors with certain metadata by entering appropriate keywords. Unfortunately, these ontology and their use are rather complex and it is uncertain 
whether end users 
can provide correct descriptions of sensors and their deployment context without the help from experts. In other words, such methods 
require extensive prior knowledge. 
There are efforts to provide a standardized vocabulary to describe sensors and their properties such as SensorML\footnote{http://www.opengeospatial.org/standards/sensorml} and the Semantic Sensor Network Ontology (SSN)\footnote{http://www.w3.org/2005/Incubator/ssn/ssnx/ssn}, but not widely adopted. 

The above solutions are time-consuming and require 
expert knowledge. For example, the descriptions of things and their corresponding characteristics and ontology need to be predefined under a uniform format such as Resource Description Framework (RDF) or Schema.org\footnote{http://schema.org/}. In addition, the methods do not make full use of the rich information on users historical interactions with things, 
which may imply containing implicit relations of different entities. For example, if some users have the similar usage pattern on certain things, it may indicate some close connections among these things. 
Existing solutions can not capture such information well. 
We propose to extract the underlying connections between things by exploiting the 
human-thing interactions in ubiquitous environment. Our method not only takes rich contextual information of human-thing interactions into account, but also utilizes the historical pattern by analyzing past human-thing interactions.

\section{Conclusion}
\label{sec:conc}

Recent advances in radio-frequency identification (RFID), wireless sensor networks, and Web services have made it 
possible to bridge the physical and digital worlds together, where ubiquitous things are becoming an integral part of our daily lives. Despite the exciting potential of this prosperous era, there are many challenges that persist. In this paper, we propose a novel model that derives latent correlations among things by exploiting user, temporal, and spatial information captured from things usage events. This correlation analysis can help solve many challenging issues in managing things such as things search, recommendation, annotation, classification and clustering. The experimental results demonstrate the utility of our approach. 

We view the work presented in this paper as a first step towards effective management of things in the 
emerging Web of Things (WoT) era. 
%
There are a few interesting directions that we plan to work in the future: 
\begin{itemize}
\item {\em Real-time things status update.}  In real situation, physical things are more dynamic compared to traditional Web resources. Examples of such dynamic features include availability, and changing attributes (e.g., geographical information, status). We plan to improve our model so that it can adaptively propagate up-to-date information from things correlations network and make more accurate recommendations.

\item {\em Scalability.} We plan to improve the scalability of our approach by adopting constraints in searching a local area. This can be realized by applying generalized clustering algorithms on hypergraphs. The search space can be significantly pruned in this way. We also plan to evaluate the improved approach using real-world large-scale WoT datasets. 
%
We notice that some parameters (e.g., $\alpha$ and $\beta$) in our approach need to be tuned in a category specific way. This might be a burden to apply the algorithm
to other WoT datasets. We will 
investigate ways to reduce the workload of parameter tuning 
e.g., by some meta-feature based methods or multi-task learning \cite{lee2007learning,zhang2011multi}.

\item {\em Thing-to-Thing communications.} Our current model works based on human-thing interactions to extract the latent connections between things. The communications between things are getting more 
prevalent with development of communication technologies, 
which represent 
a rich source to exploit for making our current model more robust. Extending our model by analyzing and exploring the thing-to-thing  communications is another future work.

%
\end{itemize}

\bibliographystyle{ACM-Reference-Format-Journals}
\bibliography{lina,linatoit}



\medskip


\end{document}